\documentclass[journal, onecolumn,12pt, draftclsnofoot]{IEEEtran}
\usepackage{citesort}
\usepackage{amsmath,amsthm}
\usepackage{algorithmic,algorithm}
\usepackage{array}
\usepackage{amsfonts}
\usepackage{graphicx,color,overpic,psfrag,epsfig}
\usepackage{bm}
\usepackage{footmisc}
\usepackage{xcolor}
\usepackage{subfigure}
\usepackage{epstopdf}
\usepackage{amssymb}
\newcommand{\qa}{{\bf a}}

    \newcommand{\qd}{{\bf d}}
    \newcommand{\qe}{{\bf e}}

    \newcommand{\qh}{{\bf h}}

    \newcommand{\ql}{{\bf l}}

    \newcommand{\qp}{{\bf p}}
    \newcommand{\qq}{{\bf q}}
    
    \newcommand{\qs}{{\bf s}}
    \newcommand{\qt}{{\bf t}}
    
    \newcommand{\qv}{{\bf v}}
    \newcommand{\qw}{{\bf w}}
    \newcommand{\qx}{{\bf x}}
    
    \newcommand{\qz}{{\bf z}}
    \newcommand{\qA}{{\bf A}}
    \newcommand{\qB}{{\bf B}}
    
    \newcommand{\qD}{{\bf D}}
    \newcommand{\qE}{{\bf E}}
    \newcommand{\qF}{{\bf F}}
    \newcommand{\qG}{{\bf G}}
    \newcommand{\qH}{{\bf H}}
    \newcommand{\qI}{{\bf I}}
    \newcommand{\qJ}{{\bf J}}

    \newcommand{\qP}{{\bf P}}

    \newcommand{\qT}{{\bf T}}
    
    \newcommand{\qV}{{\bf V}}
    \newcommand{\qW}{{\bf W}}

    \newcommand{\qzero}{{\bf 0}}
    
    \newcommand{\diag}{{\sf diag}}
    \newcommand{\tr}{{\sf tr}}
    \newcommand{\Ex}{{\sf E}}

    \newcommand{\calN}{{\mathcal N}}
    \newcommand{\calS}{{\mathcal S}}
    \newcommand{\bbC}{{\mathbb C}}
    \newcommand{\argmin}{\operatornamewithlimits{arg\, min}}
    \newcommand{\argmax}{\operatornamewithlimits{arg\, max}}
    
    \newcommand{\ntoinfty}{N\rightarrow \infty}

\newtheorem{lemma}{Lemma}
\begin{document}

\title{Power Minimization Based Joint Task Scheduling and Resource Allocation in Downlink C-RAN}

\author{Wenchao Xia\thanks{W. Xia, J. Zhang, and H. Zhu are with the Jiangsu Key Laboratory of Wireless Communications, Nanjing University of Posts and Telecommunications, Nanjing 210003, P. R. China, E-mail addresses: {\sf \{ 2015010203,zhangjun,hbz\}@njupt.edu.cn}.}, Jun Zhang, Tony Q. S. Quek\thanks{T. Q. S. Quek is with the  Information Systems Technology and Design Pillar, Singapore University of Technology and Design, Singapore 487372, E-mail address: {\sf tonyquek@sutd.edu.sg}.}, Shi Jin\thanks{S. Jin is with the National Mobile Communications Research Laboratory, Southeast University, Nanjing 210096, P. R. China, E-mail address: {\sf jinshi@seu.edu.cn}.}, and Hongbo Zhu \thanks{Parts of this work were accepted in IEEE Wireless Commun. Network Conf. (WCNC) \cite{Xia2018Energy}, Barcelona, Spain, Apr. 2018.}
}

\maketitle

\begin{abstract}
In this paper, we consider the network power minimization problem in a downlink cloud radio access network (C-RAN), taking into account the power consumed at the baseband unit (BBU) for computation and the power consumed at the remote radio heads and fronthaul links for transmission. The power minimization problem for transmission is a fast time-scale issue whereas the power minimization problem for computation is a slow time-scale issue. Therefore, the joint network power minimization problem is a mixed time-scale problem. To tackle the time-scale challenge, we introduce large system analysis to turn the original fast time-scale problem into a slow time-scale one that only depends on the statistical channel information. In addition, we propose a bound improving branch-and-bound algorithm and a combinational algorithm to find the optimal and suboptimal solutions to the power minimization problem for computation, respectively, and propose an iterative coordinate descent algorithm to find the solutions to the power minimization problem for transmission. Finally, a distributed algorithm based on hierarchical decomposition is proposed to solve the joint network power minimization problem. In summary, this work provides a framework to investigate how execution efficiency and computing capability at BBU as well as delay constraint of tasks can affect the network power minimization problem in C-RANs.
\end{abstract}

\begin{IEEEkeywords}
Cloud radio access network, large system analysis, energy efficiency, power minimization, computation resource, task scheduling.
\end{IEEEkeywords}

\section{Introduction}
During the last decade, the evolution of information and communication technology is causing energy consumption levels to reach a distressing rate, due to the dramatic increase in the quantity of subscribers and the number of  devices  \cite{gandotra2017green}. The massive connectivity also leads to tremendous carbon dioxide emissions into the environment. To reduce energy consumption, many new technologies and network architectures are proposed for 5G green communications \cite{tony2017cloud}.  Cloud radio access network (C-RAN) is a new system architecture where computational resource is aggregated into a central baseband unit (BBU) pool to implement the baseband processing of the conventional base stations. The radio functions including amplification, A/D and D/A conversion, and frequency conversion are performed at remote radio heads (RRHs) \cite{wu2015cloud}. In C-RANs, conventional  base stations are replaced with low-cost RRHs and these RRHs are deployed close to user equipment terminals (UEs), so the transmission power is significantly reduced. Furthermore, virtualization technique can take full advantage of aggregated computational resources to improve hardware unitization and centralized signal processing can achieve cooperation gain \cite{tony2017cloud,peng2015fronthaul}.

However, with the aforementioned advantages, new challenges also arise in C-RANs. With the dense deployment of RRHs, C-RANs consume considerable power. Hence turning the idle RRHs into sleep mode and designing energy efficient beamforming matrix are important issues \cite{luo2015downlink,shi2014group}.  In addition, the increased traffic causes a heavy burden on fronthaul in terms of capacity demand and power consumption  \cite{dai2016energy,Yu2016Joint}.  Finally, the power consumption of baseband processing is also considerable, which is determined by the allocation of computational resource. Overall, all the three challenges have a great effect on the network power consumption in C-RANs.

The network power minimization problem has been extensively studied in \cite{peng2014heterogeneous,Yu2016Joint,dai2016energy,luo2015downlink,shi2014group,pan2017joint}. Reference \cite{Yu2016Joint} jointly optimized downlink beamforming and admission control to minimize the network power. Reference \cite{dai2016energy} compared two transmission schemes, i.e., the data-sharing scheme and compression scheme. Reference \cite{luo2015downlink} proposed a joint downlink and uplink UE-RRH association and beamforming design to reduce energy consumption. Precoding design and RRH selection were optimized jointly in \cite{pan2017joint,shi2014group}.  However, the aforementioned papers only considered the first and second challenges, taking into account the power consumption for transmission, i.e.,  power consumptions of the RRHs and fronthaul links. Dealing with the third challenge in C-RANs is still an open issue. The computational resource aggregated in the BBU pool is provided by many physical servers. Each UE's task is first scheduled on one of these servers and then executed by a virtual machine (VM) created by the server.   Therefore, task scheduling and computational resource allocation are the key to the third challenge. There exist some works on computational resource allocation \cite{guo2016exploiting,wang2018joint,tang2015cross,liu2016computing,tang2017systematic,Tang2017System,shi2017energy}. References \cite{tang2015cross,guo2016exploiting} used a queueing model to represent UEs' data processing and transmitting behavior. Reference \cite{liu2016computing} modelled the power consumption for computation as an increasing function of UEs' rates. Reference \cite{wang2018joint} investigated a mobile cloud computing system with computational resource allocation. One thing these works have in common is that they all considered delay constraint. With the popularity of the online video and mobile game, as well as the development of the Internet of things, traffic delay is considered as a key metric to measure the quality-of-service (QoS). However, none of these works take into account task scheduling and computational resource allocation  simultaneously. Besides, these works do not consider the time-scale challenge except reference \cite{tang2017systematic}, in which the sample averaging was adopted to approximate the time averaging of the power consumption of transmission.

Motivated by these facts, we aim to minimize the network power consumption under delay constraint where the aforementioned three challenges are considered simultaneously  in this paper. We consider a downlink C-RAN composed of  many RRHs which are connected to a BBU pool via fronthaul. In the BBU pool, there is a data center with a set of physical servers. Each UE has one task which is first scheduled on a certain server and  a VM is created by the server to execute this task. Then, the output data is transmitted using RRHs  via fronthaul to the UEs. Due to limited fronthaul capacity, the precoded signals are first compressed and then the corresponding compression descriptions are forwarded through the fronthaul. We formulate a joint network power minimization problem of task scheduling and resource allocation, which includes not only computational resource allocation but also power allocation for transmission. Note that the power minimization problem for transmission is a fast time-scale issue because it depends on small-scale fading which varies in the order of milliseconds. However, the power consumption problem for computation is a slow time-scale issue since the task scheduling and computation resource allocation are usually executed much slower than milliseconds \cite{tang2017systematic}. Therefore, the joint network power minimization problem  is a mixed time-scale issue. The main contributions of this work are summarized as follows:
 \begin{itemize}
   \item We first formulate two power minimization problems for computation and transmission, respectively.  The power minimization problem for computation is a slow time-scale issue and also a mixed-integer nonlinear programming, where the task scheduling and computation resource allocation are optimized jointly. However, the power minimization problem for transmission is a  fast time-scale issue and also a nonconvex problem where power allocation and compression noise are optimized jointly. Then, a joint and mixed time-scale network power minimization problem combining the above two problems is also formulated.
   \item We translate the fast/mixed time-scale problem into a slow time-scale one. Different from reference \cite{tang2017systematic}, where the sample averaging was used to approximate the time averaging of the power consumption of transmission, we introduce the large system analysis to convert our problem into one that only depends on statistical channel information (i.e., large-scale fading) instead of small-scale fading. Therefore, the power minimization problem for transmission, as well as the joint network power minimization problem, is turned into a slow time-scale one.
   \item For the power minimization problem for computation, we propose a bound improving branch and bound (BnB) algorithm to determine the optimal solutions. To reduce the computational complexity and time, we also propose a suboptimal combinational algorithm. For  the power minimization problem for transmission, an  iterative coordinate descent algorithm is proposed to determine solutions. Finally,  a distributed algorithm based on hierarchical decomposition is proposed to solve the joint network power minimization problem.
 \end{itemize}

The remainder of this paper is organized as follows. Section \ref{section system model} introduces the system model and formulates three power  minimization problems. Section \ref{Power Minimization Problem for Computation}  proposes two algorithms, i.e., the BnB algorithm and combinational algorithm, to solve the power minimization problem for computation. Section \ref{Power Minimization  Problem for Transmission} proposes an iterative coordinate descent algorithm to solve the power minimization problem for transmission. Based on the analysis in Sections \ref{Power Minimization Problem for Computation} and  \ref{Power Minimization  Problem for Transmission}, a distributed algorithm based on hierarchical decomposition is proposed to solve the joint network power minimization problem in Section \ref{Power Minimization Problem for Computation and Transmission}. Numerical results are presented in Section \ref{section numerical results}. Finally, conclusion is drawn in Section \ref{section conclusion}.

\textbf{Notations:} The notations are given as follows. Matrices and vectors are denoted by bold capital and lowercase symbols. $(\qA)^T$, $(\qA)^\dagger$, and $\tr(\qA)$ stand for transpose, conjugate transpose, and trace of $\qA$, respectively. $\qA\succeq\qzero$ indicates that $\qA$ is a Hermitian positive semidefinite  matrix. The notations $\Ex(\bullet)$ and $||\bullet||_0$ are expectation and $l_0$ norm operators, respectively. Finally, $\qa\sim\mathcal{CN}(\qzero,\bm{\Sigma})$ is a complex Gaussian vector with zero-mean and covariance matrix $\bm{\Sigma}$.

\section{Cloud Radio Access Network}\label{section system model}
\begin{figure}
\includegraphics[width=0.8\textwidth]{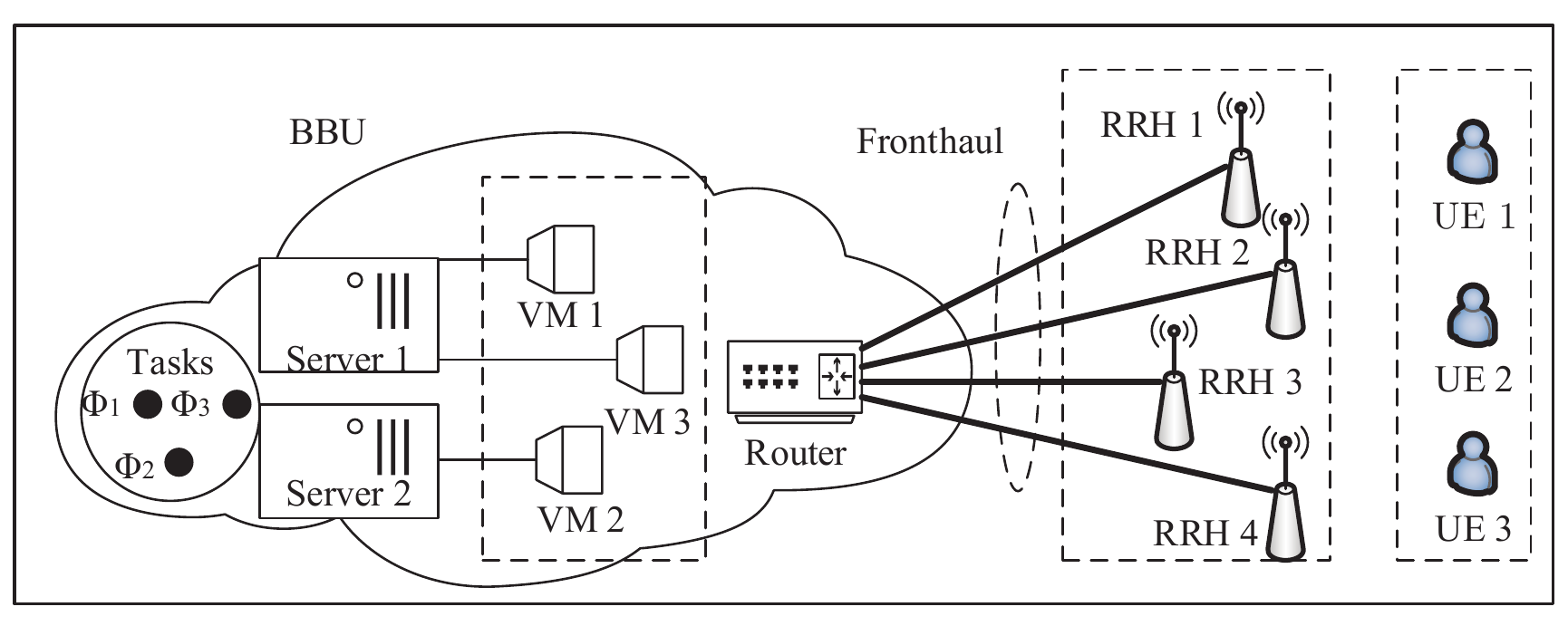}
\centering
\caption{A typical structure of downlink C-RAN with a data center.} \label{system model}
\end{figure}
\subsection{System Model}
Consider a downlink C-RAN where  $L$ RRHs, each with $N$ antennas, serve $K$ single-antenna UEs, as shown in Fig. \ref{system model}. The sets of the RRHs and UEs are denoted as $\mathcal{N}_R\triangleq\{1,2,\ldots,L\}$ and $\mathcal{N}_U\triangleq\{1,2,\ldots,K\}$, respectively. In the BBU pool, there is a data center consisting of a set of servers $\calN_S\triangleq\{1,2,\ldots,S\}$. The UEs' tasks are first processed at the data center before the output data is transmitted via the RRHs. It is assumed that the RRHs are connected to the BBU pool through high-speed but limited-capacity fronthaul links. In particular, the compress-and-forward scheme is adopted such that the signals for the UEs are first compressed and then the compression descriptions are forwarded to all the  RRHs.

In the following, we consider that each UE has one delay-sensitive and computation-intensive task to be executed at the data center. Similar to references \cite{Chen2015Decentralized,wang2018joint}, the task $\Phi_k$ of UE $k$ is modelled as
\begin{equation}
  \Phi_k=\langle D_k,\tau_k,L_k\rangle,
\end{equation}
where $\langle \cdot,\cdot,\cdot\rangle$ is a triplet, $D_k$ is the amount of output data after accomplishing task $\Phi_k$, $\tau_k$ denotes the total time constraint on task execution and data transmission, and $L_k$ represents the load of task $\Phi_k$. Here, we define the load as the execution time when it is executed on a VM with unit computation capability \cite{shi2017energy}.

The tasks are scheduled on different servers for execution at the data center. We use binary variables $x_{s,k}\in\{0,1\}$ to present the placement plan of tasks, where $x_{s,k}=1$ indicates task $\Phi_k$ is placed on server $s\in\calN_S$ and  $x_{s,k}=0$ otherwise. After the task $\Phi_k$ is placed on server $s$ with computing capacity $\lambda_s$, a VM with computing capability $A_{s,k}$ is created by server $s$ to complete task $\Phi_k$. Due to the diversity of servers, different servers have different executing efficiencies and we define $\varsigma_{s,k}$ as the efficiency of executing task $\Phi_k$ on server $s$. Note that a task can be scheduled on one and only one server during a task execution period so we have the constraint $\sum_{s\in\calN_S}{x_{s,k}}=1, \forall k\in\calN_U$. Then the corresponding execution time of task $\Phi_k$ is given as
\begin{equation}
  T^{(EX)}_k=\frac{L_k}{\sum_{s\in\calN_S}\varsigma_{s,k}x_{s,k}A_{s,k}},
\end{equation}
where $A_{s,k}$ should meet the computing capacity constraint of server $s$ as follows:
\begin{equation}
  \sum_{k\in\calN_U}{x_{s,k}A_{s,k}}\leq \lambda_s, \forall s\in\calN_S.
\end{equation}

Once one task is finished, its resulting data is encoded and forwarded to the corresponding UE. We first define the channel matrix between all UEs and RRH $l$ as $\qH_l=[\qh_{l,1},\ldots,\qh_{l,K}]\in\bbC^{N\times K}$ with $\qh_{l,k}=\sqrt{d_{l,k}}\tilde{\qh}_{l,k}$, where $d_{l,k}$ is the large-scale fading factor caused by path loss and shadow fading between UE $k$ and RRH $l$, and $\tilde{\qh}_{l,k}\sim\mathcal{CN}(\qzero,\qI_N)$ is the small-scale fading factor. We assume that the UEs are static or moving slowly such that in a task execution period the large-scale fading is invariant.

At the BBU, maximum-ratio transmission is adopted at the signal vector $\qs=[s_1,\ldots,s_K]^T\in\bbC^{K\times 1}$, where $s_k\sim\mathcal{CN}(0,1)$ is the signal for  UE $k$. The perfect channel state information is assumed to be available at the BBU, then the precoded signals for RRH $l$ is given by
\begin{equation}
  \hat{\qx}_{R_l}=\qV_{l}\qs,
\end{equation}
where $\qV_l=\xi_l\qH_l\sqrt{\qP_l}$ is the precoding matrix, $\qP_{l}\in\bbC^{K\times K}$ is a diagonal matrix whose elements are adjustable such that power allocation is implemented to improve system performance, and $\xi_l$ is the power scale factor which is given as $\xi^2_l=\frac{1}{\Ex[\tr(\qH_l\qH_l^\dagger)]}$.

Due to the limited capacities of fronthaul links, the precoded signal $ \hat{\qx}_{R_l}$ are first independently compressed and transmitted to the RRHs via fronthaul links. Here, we adopt point-to-point (P2P) compression for simplicity\footnote{There are two common compress-and-forward schemes, i.e., P2P compression  and Wyner-Ziv (WZ) coding. WZ coding can achieve higher performance and make better use of limited fronthaul capacity than P2P compression scheme. However, such benefits come with a cost in terms of computational complexity. Besides, finding an optimal decompression order is a hard problem. In this work, for simplicity, we only consider P2P compression scheme. However, this work can be extended to the case of WZ coding scheme applied at fronthaul with a fixed decompression order.}. The quantized signal is expressed as
\begin{equation}
  \qx_{R_l}=\hat{\qx}_{R_l}+\qq_l,
\end{equation}
where $\qq_l\sim\mathcal{CN}(0,\bm{\Psi}_l)$ is the quantization noise independent of signal $\hat{\qx}_{R_l}$ with $\bm{\Psi}_l\triangleq \Ex(\qq_l\qq_l^{\dag})$. Note that the process of signal compression is independent so that the quantization noise signals $\qq_l$ and $\qq_{l^{\prime}}$, are uncorrelated, i.e., $\Ex(\qq_l\qq_{l^{\prime}}^{\dag})=\qzero, l^{\prime}\neq l$.  According to reference \cite{el2011network}, the signal $\qx_{R_l}$ can be recovered from $\hat{\qx}_{R_l}$  at  RRH $l$ if the condition
\begin{align}
 R_{F_l}&=\Ex\left(\log_2\frac{|\qV_{l}\qV_{l}^{\dag}+\bm{\Psi}_l|}{|\bm{   \Psi}_l|}\right)\leq C_l, l\in\calN_R,\label{power constraint}
\end{align}
is satisfied, where $C_l$ is the fronthaul capacity  for RRH $l$. Furthermore, the transmission power at RRH $l$ should meet the power constraint given as follows:
\begin{equation}
  P_{R_l}^{(TR)}=\Ex\left[\tr\left(\qV_{l}\qV_{l}^\dag+\bm{\Psi}_l\right)\right]\leq P^{(MAX)}_{R_l}, \forall l\in\mathcal{N}_R,
\end{equation}
where $P^{(MAX)}_{R_l}$ is the transmission power budget.

The received signal at UE $k$ is given by
\begin{equation}
  y_{U_k}=\qh^\dag_{k}\qx_R+z_{U_k},
\end{equation}
where $\qh_k=[\qh_{1,k}^T,\dots,\qh_{L,k}^T]^T\in\bbC^{LN\times 1}$, $\qx_R=[\qx^T_{R_1},\ldots,\qx^T_{R_L}]^T\in\bbC^{LN\times 1}$, and $z_{U_k}\sim\mathcal{CN}(0,\sigma^2)$ is the independent received noise with zero mean and variance $\sigma^2$. Although the UEs do not know the exact effective channels, we assume that the average effective channels can be learned at the UEs. Therefore, the achievable rate of UE $k$ using a standard bound based on the worst-case uncorrelated additive noise \cite{hassibi2003much,hoydis2013massive}  is computed as
\begin{equation}
  R_{U_k}=B\log_2\left(1+\frac{\text{Sig}_k}{\text{Int}_k}\right),
\end{equation}
where $B$ is the system bandwidth, $\text{Sig}_k=|\Ex(\qh^\dag_{k}\qv_{k})|^2$, $\qv_{k}=[\qv_{1,k}^T,\dots,\qv_{L,k}^T]^T\in\bbC^{LN\times 1}$, and
\begin{equation}
  \text{Int}_k=\text{var}(\qh^\dag_{k}\qv_{k})+\sum\limits_{i\in\calN_U\setminus\{k\}}\Ex|{\qh_{k}^\dag\qv_{i}|^2+\Ex|\qh^\dag_{k}\bm{\Psi}\qh_{k}|}+\sigma^2,
\end{equation}
where $\text{var}(x)=\Ex\{[x-\Ex(x)][x-\Ex(x)]^\dagger\}$ and $\bm{\Psi}=\text{diag}(\{\bm{\Psi}_l\}_{l=1}^L)$. Then, the transmission time\footnote{In this paper, we assume that the transmission time at fronthaul links is constant and negligible so that it  can  be ignored.} of output data of task $\Phi_k$ is given as $T^{(TR)}_{k}=\frac{D_k}{R_{U_k}}$.

\subsection{Power Consumption Model}
In the following, we are interested in the network power  which includes the powers consumed at the RRHs, the fronthaul links, and the servers.
\subsubsection{RRH Power Consumption}
The power consumption at the RRHs consists of both circuit power consumption and transmitting power consumption,  and we adopt a linear power consumption model given by \cite{shi2014group,Auer2011how}
\begin{equation}
P_{R_l}=
\begin{cases}
\frac{1}{\upsilon_l}P^{(TR)}_{R_l}+P^{(Active)}_{R_l}, \ &\text{if}\ P^{(TR)}_{R_l}>0,\\
P^{(Sleep)}_{R_l},\ &\text{if} \ P^{(TR)}_{R_l}=0,
\end{cases}
\end{equation}
where $\upsilon_l$ is the efficiency of the power amplifier and $P^{(Active)}_{R_l}$ denotes the circuit power consumption to support RRH $l$ to transmit signals. If there is no transmission at RRH $l$, it can be turned into sleep mode with lower power consumption $P^{(Sleep)}_{R_l}$. Generally, $P^{(Sleep)}_{R_l}<P^{(Active)}_{R_l}$, thus turning a RRH into sleep mode can save power. We define $P_{\triangle R_l}=P^{(Active)}_{R_l}-P^{(Sleep)}_{R_l}$ and $P_{R_l}$ can be rewritten as
\begin{equation}
P_{R_l}=\frac{1}{\upsilon_l}P^{(TR)}_{R_l}+P^{(Sleep)}_{R_l}+|| P^{(TR)}_{R_l}||_0 P_{\triangle R_l}.
\end{equation}

\subsubsection{Fronthaul Power Consumption}
The power consumption model of fronthaul links depends on specific fronthaul technologies. Similar to reference \cite{dai2016energy}, we use a general model to compute the power consumption of each fronthaul channel as
\begin{equation}
  P_{F_l}=\eta_l R_{F_l},
\end{equation}
where $\eta_l=\frac{P^{(MAX)}_{F_l}}{C_l}$ and $P^{(MAX)}_{F_l}$ is the power consumed by the fronthaul link for RRH $l$ when working at full capacity. This model has been used for microwave backhaul links in \cite{fehske2010bit} and also can be generalized to other backhaul technologies, such as passive
optical network, fiber-based Ethernet, etc., as mentioned in \cite{Wu2016Green}.

\subsubsection{Server Power Consumption}
The total power consumption of a  server $s$ is given by \cite{Husain2010VMeter}
\begin{equation}
  P_{S_s}=P^{(Static)}_{S_s}+\sum_{k\in\calN_U}P_{VM_{s,k}},
\end{equation}
where $P^{(Static)}_{S_s}$ is  constant no matter whether VMs are running or not and  $P_{VM_{s,k}}$ is the power consumed by a VM. It is  observed that the total power consumption of a VM is directly related to the system component utilization \cite{Husain2010VMeter,shi2017energy,beloglazov2012energy}. More utilization of the system components leads to more power consumption \cite{Husain2010VMeter}. In the linear weighted model, the total power consumption $P_{VM_{s,k}}$ of VM $k$ created by server $s$ can be further decomposed into four components related to CPU, disk, IO devices, and memory as follows \cite{shi2017energy}:
\begin{equation}
  P_{VM_{s,k}}=P^{(CPU)}_{VM_{s,k}}+P^{(DISK)}_{VM_{s,k}}+P^{(IO)}_{VM_{s,k}}+P^{(MEMORY)}_{VM_{s,k}}.
\end{equation}
Because there exists a direct relation between the execution time of tasks on VMs and CPU utilization, we use the CPU power consumption to approximate the VM power consumption with a weight $\chi_{s,k}$ \cite{beloglazov2012energy,shi2017energy}, which can be expressed as
\begin{equation}
  P_{VM_{s,k}}=x_{s,k}\chi_{s,k} A_{s,k}.
\end{equation}

\subsection{Problem Formulation}
Since we are interested to minimize the network power consumption while meeting the delay constraint, we first formulate two  power minimization problems for computation and transmission, respectively. Then,  a joint network power minimization problem is also established.
\subsubsection{Power Minimization  Problem for Computation}
It is assumed that the time limitation for finishing task $\Phi_k$ on a certain VM is $\tau^{(EX)}_k$ ($\tau^{(EX)}_k<\tau_k$). Based on the above analysis, the power minimization problem for computation where task scheduling and computational resource allocation are executed jointly is formulated as
\begin{subequations}
\begin{align}
  \mathcal{P}_{0}:\min_{\qx,\qA}  \ \ &\sum_{s\in\calN_S}P_{S_s} \\
  \text{s.t.}\ \ &T^{(EX)}_k\leq \tau^{(EX)}_{k}, \forall k\in\mathcal{N}_U, \label{P0 cond1}\\
  &\sum_{k\in\calN_U}{x_{s,k}A_{s,k}}\leq \lambda_s,\forall s\in\calN_S,\label{P0 cond2}\\
  &\sum_{s\in\calN_S}{x_{s,k}}=1, \forall k\in\calN_U,  \label{P0 cond3}\\
  &A_{s,k}\geq 0, \forall k\in\mathcal{N}_U, \forall s\in\calN_S, \label{P0 cond4}\\
  &x_{s,k}\in \{0,1\}, \forall k\in\mathcal{N}_U, \forall s\in\calN_S. \label{P0 cond5}
\end{align}
\end{subequations}
where  $\qx$ is a collect of $x_{s,k}$'s, indicating the placement plan of tasks and  $\qA$ is a collect of $A_{s,k}$'s denoting the resource allocation plan of the servers.
\subsubsection{Power Minimization  Problem for Transmission}
Similarly, we first assume that the time constraint for transmitting the output signals of the tasks is $\tau^{(TR)}_k$ ($\tau^{(TR)}_k<\tau_k$). Then, we formulate the power minimization problem for transmission as
\begin{subequations}
\begin{align}
\mathcal{P}_{1}:\min_{\qP,\bm{\Psi}}  \ \ &\sum_{l\in\calN_R}P_{R_l}+P_{F_l} \\
  \text{s.t.}\ \ &T^{(TR)}_k\leq \tau^{(TR)}_{k}, \forall l\in\mathcal{N}_R,  \label{P1 cond1}\\
  &R_{F_l}\leq C_l, \forall l\in\mathcal{N}_R, \label{P1 cond2}\\
  &P_{R_l}^{(TR)}\leq P^{(MAX)}_{R_l}, \forall l\in\mathcal{N}_R, \label{P1 cond3}
\end{align}
\end{subequations}
where $\qP$ is a collect of $p_{l,k}$'s and  $\bm{\Psi}$ is a collect of $\bm{\Psi}_l$'s. Note that $\qP$ describes the power allocation scheme and $\bm{\Psi}$ indicates the quantization levels of the all RRHs.
\subsubsection{Joint Network Power Minimization Problem}
Finally, the joint network power minimization problem for computation and transmission is formulated as
\begin{subequations}\label{problem2}
\begin{align}
  \mathcal{P}_{2}:\min_{\qx,\qA,\qP,\bm{\Psi}}  \ \ &\sum_{s\in\calN_S}P_{S_s}+\omega\sum_{l\in\calN_R}(P_{R_l}+P_{F_l})\\
  \text{s.t.}\ \ &T^{(EX)}_k+T^{(TR)}_k\leq \tau_{k}, \forall k\in\mathcal{N}_U, \\
  &\eqref{P0 cond2}-\eqref{P0 cond5},\eqref{P1 cond2}, \ \text{and} \ \eqref{P1 cond3},
\end{align}
\end{subequations}
where $\omega$ is a factor to balance the power consumption of computation and transmission.

We observe that $\mathcal{P}_{0}$ is a slow time-scale problem but the joint optimization of power allocation and quantization noise in  $\mathcal{P}_{1}$ is a fast time-scale problem since it depends on small-scale fading. Consequently,  $\mathcal{P}_{2}$ is a mixed time-scale issue that needs further attention \cite{tang2017systematic}. To solve this challenge caused by the time-scale issue, authors in \cite{tang2017systematic} used  ensemble averaging over fast time-scale samples  so that the final problem became a slow time-scale problem.  Instead, we  introduce large system analysis to transform $\mathcal{P}_{1}$ and $\mathcal{P}_{2}$ into slow time-scale problems  depending only on  large-scale fading \cite{zhang2013large,xia2017large}. Furthermore, we assume that the UEs are static or moving slowly such that the large-scale fading remains  invariant within a task execution period.

\section{Power Minimization  Problem for Computation}\label{Power Minimization Problem for Computation}
For $\mathcal{P}_{0}$ to be solvable, it is assumed that task $\Phi_k$ can be further divided into $S$ sub-task $\phi_{s,k}$'s, each  with load $l_{s,k}$, and placed on $S$ servers, respectively \cite{bharadwaj1996scheduling,shi2017energy}. This assumption can be interpreted as a relaxation of the binary variable $x_{s,k}$ to a real variable, i.e., $x_{s,k}\in[0,1]$, then the variable $x_{s,k}$ is absorbed in the new defined variable $l_{s,k}=x_{s,k}L_k$. The total load of sub-tasks should satisfy the constraint $\sum_{s\in\calN_S}{l_{s,k}}=L_k$. Then, a VM with computation capability $a_{s,k}$ is created by server $s$ for sub-task $\phi_{s,k}$ and the associated execution time is $t^{(EX)}_{s,k}=\frac{l_{s,k}}{\varsigma_{s,k}a_{s,k}}$, where $a_{s,k}$ satisfies the constraint $\sum_{k\in\calN_U}a_{s,k}\leq \lambda_s$. Accordingly, the power consumption of sub-task $\phi_{s,k}$ is given as $p_{VM_{s,k}}=\chi_{s,k}a_{s,k}$ and the relaxed version of $\mathcal{P}_{0}$ can be written as
\begin{subequations}
\begin{align}
  \mathcal{P}_{0\text{-}1}:\min_{\qa,\ql}  \ \ &\sum_{s\in\calN_S}\sum_{k\in\calN_U}p_{VM_{s,k}} \\
  &l_{s,k}-\varsigma_{s,k}a_{s,k}\tau^{(EX)}_k\leq 0, k\in\calN_U, s\in\calN_S,\label{subcond1}\\
  &\sum_{k\in\calN_U}a_{s,k}\leq \lambda_s, s\in\calN_S,\label{subcond2}\\
  &\sum_{s\in\calN_S}{l_{s,k}}=L_k, k\in\calN_U,\label{subcond3}\\
  &a_{s,k}\geq 0,l_{s,k}\geq 0, k\in\calN_U, s\in\calN_S\label{subcond4},
\end{align}
\end{subequations}
where $\qa$ is a collect of $a_{s,k}$'s and $\ql$ is a collect of $l_{s,k}$'s. Note that $P^{(Static)}_{S_s}$ is constant and thus omitted. The objective function and constraints \eqref{subcond1}-\eqref{subcond4} are linear so $\mathcal{P}_{0\text{-}1}$ can be solved  easily. However, the solution determined from $\mathcal{P}_{0\text{-}1}$ is generally not the optimal solution to $\mathcal{P}_{0}$. In what follows, we introduce the BnB algorithm to find the optimal solution to  $\mathcal{P}_{0}$ based on the solution to $\mathcal{P}_{0\text{-}1}$.
\subsection{Branch and Bound Algorithm}
We define a set $\calS=\{(s,k)|\forall s\in\calN_S,\forall k\in\calN_U\}$ that contains all the task-server pairs and introduce another two task-server pair sets $\calS_0=\{(s,k)|x_{s,k}=0,\forall s\in\calN_S,\forall k\in\calN_U\}$ and $\calS_1=\{(s,k)|x_{s,k}=1,\forall s\in\calN_S,\forall k\in\calN_U\}$. With the defined sets, we formulate an equivalent problem of $\mathcal{P}_{0}$ as follows:
\begin{subequations}
\begin{align}
  \mathcal{P}_{0\text{-}2}:\min_{\qx,\qA}  \ \ &\sum_{s\in\calN_S}\sum_{k\in\calN_U}P_{VM_{s,k}} \\
  \text{s.t.}\ \ &\eqref{P0 cond1}-\eqref{P0 cond4},\\
  &x_{s,k}=1,\forall (s,k)\in\calS_1,\label{p03 cond1}\\
  &x_{s,k}=0,\forall (s,k)\in\calS_0,\label{p03 cond2}\\
  &x_{s,k}\in\{0,1\}, (s,k)\in\calS\setminus(\calS_0\cup\calS_1).
\end{align}
\end{subequations}
Similarly, an equivalent problem of  $\mathcal{P}_{0\text{-}1}$ is formulated as
\begin{subequations}
\begin{align}
  \mathcal{P}_{0\text{-}3}:\min_{\qa,\ql}  \ \ &\sum_{s\in\calN_S}\sum_{k\in\calN_U}p_{VM_{s,k}} \\
  \text{s.t.}\ \ &\eqref{subcond1}-\eqref{subcond4},\\
  &l_{s,k}=L_k, \forall (s,k)\in\calS_1,\\
  &l_{s,k}=0,\forall (s,k)\in\calS_0,\\
  &0\leq l_{s,k}\leq L_k, (s,k)\in\calS\setminus(\calS_0\cup\calS_1).
\end{align}
\end{subequations}
For notational convenience, we use the related parameter tuples $(z,\calS_0,\calS_1)$ and $(z,\calS_0,\calS_1)^{\prime}$ to denote $\mathcal{P}_{0\text{-}2}$ and $\mathcal{P}_{0\text{-}3}$, respectively, where $z$ is the optimal value of the objective function in  $\mathcal{P}_{0\text{-}3}$.
The BnB algorithm for  $\mathcal{P}_{0}$ is provided in \textbf{Algorithm \ref{BnB algorithm}}. At the beginning, we define $\digamma$ as the set of branch problems and $z^{\star}$ as the best-known objective value. The main process of the BnB algorithm consists of two important steps as follows:

1) \textbf{Branching:} In each iteration process, we choose the problem that achieves the minimum lower bound, denoted as $(\hat{z},\hat{\calS}_0,\hat{\calS}_1)$, to branch. Then, the task-server pair with the highest priority $(s^{\ast},k^{\ast})$ is chosen to be divided into two smaller branch problems: one is with $x_{s^{\ast},k^{\ast}}=0$ and the other is with $x_{s^{\ast},k^{\ast}}=1$. Accordingly, the relaxed problems of the two branches are given as: one is with $l_{s^{\ast}, k^{\ast}}=0$ and the other is with $l_{s^{\ast}, k^{\ast}}=L_{k^{\ast}}$. Evidently, the priority function plays an important role in reducing the complexity and we define the priority function as $f_p(s,k)=\frac{\chi_{s,k}L_k}{\varsigma_{s,k}}$.

2) \textbf{Bounding and Pruning:} According to the selected branch, we  compute the lower bounds of sub-problems $(z^{(B1,n)},\calS^{(B1,n)}_0,\calS^{(B1,n)}_1)^{\prime}$ and  $(z^{(B2,n)},\calS^{(B2,n)}_0,\calS^{(B2,n)}_1)^{\prime}$, respectively. The two branch problems are stored in $\digamma$ for further branching when their lower bounds are less than the current best-known value $z^{\star}$. If a new feasible solution is found which is lower than the current best-known value $z^{\star}$, the current best-known solution is updated. Besides, the stored branches in $\digamma$ having an lower bound larger than the value of the new best-known feasible solution can be deleted.

\begin{algorithm}[htb]
\caption{BnB algorithm for task scheduling.}
\label{BnB algorithm}
\begin{algorithmic}[1]
\STATE \textbf{Initialization:} $z^{\star}=+\infty$, $\calS^{(0)}_0=\calS^{(0)}_1=\emptyset$, and $\digamma=\{(z^{(0)},\calS^{(0)}_0,\calS^{(0)}_1)\}$, and $n=0$. \\
\WHILE {$\digamma\neq\emptyset$}
\STATE Find the problem $(\hat{z},\hat{\calS}_0,\hat{\calS}_1)$ according to $\hat{z}=\min_{(z,\calS_0,\calS_1)\in\digamma}z$ from $\digamma$ and update $\digamma=\digamma\setminus\{(\hat{z},\hat{\calS}_0,\hat{\calS}_1)\}$.
\STATE Select the task-server pair with the highest priority, i.e., $(s^{\ast},k^{\ast})=\argmax_{(s,k)\in\calS\setminus(\hat{\calS}_0\cup\hat{\calS}_1)}f_p(s,k)$, and set $n=n+1$.
\STATE Update $\calS^{(B_1,n)}_0=\hat{\calS}_0\cup\{(s^{\ast},k^{\ast})\}$, $\calS^{(B_1,n)}_1=\hat{\calS}_1$, $\calS^{(B_2,n)}_0=\hat{\calS}_0$, and $\calS^{(B_2,n)}_1=\hat{\calS}_1\cup\{(s^{\ast},k^{\ast})\}$;
\STATE Solve problems $(z^{(B_i,n)}, \calS^{(B_i,n)}_0,\calS^{(B_i,n)}_1)^{\prime},i=1,2$. If there is no feasible solution, set $z^{(B_i,n)}=+\infty$.
\IF {$z^{(B_i,n)}<z^{\star},i=1,2$,}
\IF {$\calS==\calS^{(B_i,n)}_0\cup\calS^{(B_i,n)}_1$,}
\STATE Set $z^{\star}=z^{(B_i,n)}$, $\calS^{\star}_0=\calS^{(B_i,n)}_0$, and $\calS^{\star}_1=\calS^{(B_i,n)}_1$.
\ELSE
\STATE Update $\digamma=\digamma\cup\{(z^{(B_i,n)}, \calS^{(B_i,n)}_0,\calS^{(B_i,n)}_1)\}$.
\ENDIF
\ENDIF
\STATE Check and prune existing branches.  If branch problem $(z^{(j)}, \calS^{(j)}_0,\calS^{(j)}_1)$ in $\digamma$ meets the constraint $z^{(j)}>z^{\star}$, $j=1,2,\ldots,|\digamma|$, then it can be pruned, i.e., $\digamma=\digamma\setminus\{(z^{(j)}, \calS^{(j)}_0,\calS^{(j)}_1)\}$.
\ENDWHILE
\STATE Return $z^{\star}$, $\calS^{\star}_0$, and $\calS^{\star}_1$.
\end{algorithmic}
\end{algorithm}

\subsection{Suboptimal  Task Scheduling Algorithm}

Although the BnB algorithm can find the global optimal solution, the convergence rate can be  slow, especially for large number task-server pairs. Therefore, we introduce a suboptimal but fast task scheduling algorithm which is referred to as heuristic task scheduling algorithm, as shown in \textbf{Algorithm \ref{heuristic algorithm}}.

In \textbf{Algorithm \ref{heuristic algorithm}},  the unscheduled task $\Phi_{k^{\ast}}$ with the highest load is first considered and server $s^{\ast}$ which has the highest execution efficiency for this task has a priority. When the available resource in server $s^{\ast}$ is sufficient to support task $\Phi_{k^{\ast}}$, then server $s^{\ast}$ allocates as little computing resource as possible to task $\Phi_{k^{\ast}}$, i.e., $A_{s^{\ast},k^{\ast}}=\frac{L_{k^{\ast}}}{\tau_{k^{\ast}}\varsigma_{s^{\ast},k^{\ast}}}$. Otherwise, task $\Phi_k^{\ast}$ continues to search the potential server. Note that different from the BnB algorithm, the heuristic task scheduling algorithm cannot always find solutions to $\mathcal{P}_{0}$. However, in the case with high execution efficiency or abundant computation resource, \textbf{Algorithm \ref{heuristic algorithm}} can achieve satisfying performance with lower computational complexity and time. Therefore, we propose a combinational algorithm where \textbf{Algorithm \ref{heuristic algorithm}} is first adopted to find the suboptimal solutions.  If no solution is found via \textbf{Algorithm \ref{heuristic algorithm}},  we  continue to resort to \textbf{Algorithm \ref{BnB algorithm}}. We refer to such an algorithm as combinational task scheduling algorithm, as shown in \textbf{Algorithm \ref{combinational algorithm}}.

\begin{algorithm}[htb]
\caption{Heuristic task scheduling algorithm.}
\label{heuristic algorithm}
\begin{algorithmic}[1]
\STATE \textbf{Initialize} $\calN_{S^{\prime}}=\calN_S$.  Find task $\Phi_{k^{\ast}}=\argmax_{k\in\calN_U}{L_k}$ and update $\calN_U\triangleq\calN_U\setminus\{k^{\ast}\}$.
\IF {$\calN_{S^{\prime}}$ is not empty,}
\STATE \ \  Find server $s^{\ast}=\argmin_{s\in\calN_{S^{\prime}}}{\frac{\chi_{s,k}}{\varsigma_{s,k^{\ast}}}}$ for task $\Phi_{k^{\ast}}$ and update $\calN_{S^{\prime}}=\calN_{S^{\prime}}\setminus\{s^{\ast}\}$.
\IF  {$\frac{L_{k^{\ast}}}{\tau_{k^{\ast}}\varsigma_{s^{\ast},k^{\ast}}}\leq \lambda_{s^{\ast}}$,}
\STATE Update $A_{s^{\ast},k^{\ast}}=\frac{L_{k^{\ast}}}{\tau_{k^{\ast}}\varsigma_{s^{\ast},k^{\ast}}}$ and $\lambda_{s^{\ast}}=\lambda_{s^{\ast}}-A_{s^{\ast},k^{\ast}}$. Then, go to step 1.
\ELSE
\STATE Go to step 2.
\ENDIF
\ELSE
\STATE  Heuristic task scheduling fail.
\ENDIF
\end{algorithmic}
\end{algorithm}

\begin{algorithm}[htb]
\caption{Combinational task scheduling algorithm.}
\label{combinational algorithm}
\begin{algorithmic}[1]
\STATE \textbf{Algorithm \ref{heuristic algorithm}} is adopted.
\IF {no available solution is found via \textbf{Algorithm \ref{heuristic algorithm}},}
\STATE \textbf{Algorithm \ref{BnB algorithm}} is adopted to find the optimal solution.
\ELSE
\STATE Return the solution found by \textbf{Algorithm \ref{heuristic algorithm}}.
\ENDIF
\end{algorithmic}
\end{algorithm}

\section{Power Minimization  Problem for Transmission}\label{Power Minimization  Problem for Transmission}
In this section, we first introduce approximate results with large system analysis and then find  the solution to  $\mathcal{P}_{1}$ based on these approximations. According to large system analysis, we can take care of  the small-scale fading  using the following lemma.

\begin{lemma}
Given that $\tilde{\qh}_{l,k}$'s are i.i.d. complex Gaussian variables with independent real and imaginary parts. According to the law of large numbers and the large-dimensional random matrix theory, as $\ntoinfty$, then we have the following results:

1) $P^{(TR)}_{R_l}-\bar{P}^{(TR)}_{R_l}\rightarrow0$, where
\begin{equation}
  \bar{P}^{(TR)}_{R_l}=\bar{\xi}^2_lN\sum\limits_{k\in\calN_U}{p_{l,k}d_{l,k}}+\tr(\bm{\Psi}_l),\label{appro 1}
\end{equation}
with $\bar{\xi}^2_l=\frac{1}{N\sum_{k\in\calN_U}d_{l,k}}$.

2) $R_{U_k}-\bar{R}_{U_k}\rightarrow0$, where
\begin{align}
  \bar{R}_{U_k}&=\log_2\left[1+\frac{ \overline{\text{Sig}}_k }{\overline{\text{Int}}_k}\right], \label{appro 2}
\end{align}
where $\overline{\text{Sig}}_k=(\bar{\qd}^{T}_{k} \sqrt{\qp_{k}})^2$, $\overline{\text{Int}}_k=\frac{1}{N}\sum_{i\in\calN_U\setminus\{k\}} (\bar{\qd}_{i}\circ\bar{\qd}_{k})^T\qp_{i}+\frac{1}{N^2}\sum_{l\in\calN_R}{d_{l,k}\tr(\bm{\Psi}_l)}+\frac{1}{N^2}\sigma^2$, $\bar{\qd}_{k}=[\xi_1d_{1,k},\ldots,\xi_Ld_{L,k}]^T$, $\qp_k=[p_{1,k},\ldots,p_{L,k}]^T$, and $\bar{\qd}_{i}\circ\bar{\qd}_{k}$ is the Hadamard product whose $l$-th element is $\xi^2_ld_{l,i}d_{l,k}$.

3) $R_{F_l}-\bar{R}_{F_l}\rightarrow0$, where
\begin{equation}
  \bar{R}_{F_l}=\frac{1}{\log2}(\Delta_l-\log|\bm{\Psi}_l|),\label{appro 3}
\end{equation}
where $\Delta_l=\log|\bm{\Lambda}_l|+\sum_{k\in\calN_U}(\frac{1}{1+e_{l,k}}-\log\frac{1}{1+e_{l,k}})-K$, $e_{l,k}=\bar{\xi}^2_lp_{l,k}d_{l,k}\tr\bm{\Lambda}_l^{-1}$,  and
\begin{equation}
 \bm{\Lambda}_l=\sum_{k\in\calN_U}\frac{\bar{\xi}^2_lp_{l,k}d_{l,k}}{1+e_{l,k}}\qI_N+\bm{\Psi}_l.
\end{equation}
\end{lemma}
\IEEEproof See the Appendix.

The approximate results in \eqref{appro 1}, \eqref{appro 2}, and \eqref{appro 3}  are obtained with the assumption that $N\rightarrow\infty$. Note that these results can achieve  satisfying accuracy even when $N$ is not too large.

Based on the approximate results, we formulate an alternative to $\mathcal{P}_{1}$ as:
\begin{subequations}
\begin{align}
  \mathcal{P}_{1\text{-}1}:\min_{\qP,\bm{\Psi}}  \ \ &\sum_{l\in\calN_R}\bar{P}_{R_l}+\bar{P}_{F_l} \\
  \text{s.t.}\ \ &(2^{\frac{D_k}{\tau^{(TR)}_{k}B}}-1)\overline{\text{Int}}_k\leq \overline{\text{Sig}}_k, \forall k\in\mathcal{N}_U, \label{P11 cond1}\\
  &\bar{R}_{F_l}\leq C_l, \forall l\in\mathcal{N}_R,\label{P11 cond2} \\
  &\bar{P}^{(TR)}_{R_l}\leq P^{(MAX)}_{R_l}, \forall l\in\mathcal{N}_R, \label{P11 cond3}
\end{align}
\end{subequations}
where $\bar{P}_{F_l}=\eta_l\bar{R}_{F_l}$ and $\bar{P}_{R_l}=(\frac{1}{\upsilon_l}+\rho_lP_{\triangle R_l})\bar{P}^{(TR)}_{R_l}+P^{(Sleep)}_{R_l}$. According to reference \cite{Candes2008}, the \textit{$l_0$}-norm can be approximated with convex relaxation \textit{$l_1$}-norm as $|| \bar{P}^{(TR)}_{R_l}||_0 \approx  \rho_l \bar{P}^{(TR)}_{R_l}$, where $\rho_l=\frac{c_1}{\bar{P}^{(TR)}_{R_l}+c_2}$ is iteratively updated, $c_1$ is a constant, and $c_2$ is a small constant to guarantee numerical satiability.  However, $\mathcal{P}_{1\text{-}1}$ is still non-convex with respect to $p_{l,k}$ and $\bm{\Psi}_l$ because of $\bar{R}_{F_l}$ and $\overline{\text{Sig}}_k$. To achieve a stationary point of  $\mathcal{P}_{1\text{-}1}$, we first introduce the following lemma.

\begin{lemma}[\cite{li2013transmit,zhou2014optimized}]\label{lemma1}
For any two $N\times N$  positive  definite Hermitian matrices $\bm{\Lambda}$ and $\bm{\Gamma}$, then
\begin{equation}\label{inequality1}
  \log|\bm{\Lambda}|\leq-\log|\bm{\Gamma}|+\tr(\bm{\Gamma}\bm{\Lambda})-N,
\end{equation}
with the equality if and only if $\bm{\Gamma}=\bm{\Lambda}^{-1}$.When $N=1$, the inequality \eqref{inequality1} is simplified as
\begin{equation}\label{inequality2}
  \log(\varphi)\leq-\log(\gamma)+\varphi\gamma-1,
\end{equation}
with the equality if and only if $\gamma=\varphi^{-1}$.
\end{lemma}
Applying  \eqref{inequality1} to the denominator of $\bar{R}_{F_l}$, then we have
\begin{equation}
  \tilde{\bar{R}}_{F_l}=-\log_2|\bm{\Gamma}_{l}|+\tr(\bm{\Gamma}_{l}\bm{\Lambda}_l)-N-\log_2|\bm{\Psi}_l|+\sum_{k\in\calN_U}(\frac{1}{1+e_{l,k}}-\log\frac{1}{1+e_{l,k}})-K,
\end{equation}
which is equivalent to $ \bar{R}_{F_l}$ when $\bm{\Gamma}_{l}=\bm{\Lambda}^{-1}_l$.

Next, we change optimization variable $p_{l,k}$ to $\qW=\qw\qw^T\in\bbC^{KL\times KL}$ where $\qw=[\qw^T_1,\ldots,\qw^T_K]^T$ and $\qw_k=\sqrt{\qp_k}$. Then, $\bar{P}^{(TR)}_{R_l}$ can be rewritten as
\begin{equation*}
\bar{P}^{(TR)}_{R_l}=N\tr(\qA_l\qT\qW)+\tr(\qB_l\bm{\Psi}),
\end{equation*}
where $\qA_l=\text{diag}([\qa^T_l,\ldots,\qa^T_l]^T)\in\bbC^{KL\times KL}$, $\qa_{l}\in\bbC^{L\times 1}$ represents a vector whose  $l$-th element is 1 and  0 elsewhere, $\qT=\text{diag}(\{\qt_k\}_{k=1}^K)\in\bbC^{KL\times KL}$ is a diagonal matrix, $\qt_k\in\bbC^{L\times 1}$ denotes a vector whose $l$-th element is $\bar{\xi}^2_ld_{l,k}$, and $\qB_l\in\bbC^{NL\times NL}$ is a diagonal matrix whose main diagonal elements from $((l-1)N+1)$-th to $(lN)$-th are 1's and  0 elsewhere.  $\bm{\Lambda}_l$ can be rewritten as
\begin{equation*}
\bm{\Lambda}_l=\tr(\qA_l\qE\qA_l\qW)\qI_N+\qJ_l\qB_l\bm{\Psi}\qJ_l^H,
\end{equation*}
where  $\qE=\text{diag}(\{\qe_k\}_{k=1}^K)\in\bbC^{KL\times KL}$ is a diagonal matrix, $\qe_k$ is a vector whose $l$-th  diagonal element is $\frac{\bar{\xi}^2_ld_{l,k}}{1+e_{l,k}}$, and $\qJ=[\qzero_1,\ldots,\qzero_{l-1},\qI_N,\qzero_{l+1}\ldots,\qzero_L]\in\bbC^{N\times NL}$ with $\qzero_l\in\bbC^{N\times N}$ being a zero-matrix. Similarly,  based on the new defined variable $\qW$, $e_{l,k}$, $\overline{\text{Sig}}_k$, and $\overline{\text{Int}}_k$ can be rewritten as $e_{l,k}=\tr(\qT\qG_{lk}\qW)\tr\bm{\Lambda}_l^{-1}$, $\overline{\text{Sig}}_k=\tr(\qF_k\bar{\qD}\qF_k\qW)$,
and
\begin{equation*}
\overline{\text{Int}}_k=\frac{1}{N}\sum_{i\in\calN_U\setminus\{k\}}\tr(\tilde{\qD}_{ik}\qF_i\qW)+\frac{1}{N^2}\tr(\qD_k\bm{\Psi})+\frac{1}{N^2}\sigma^2, \end{equation*}
respectively, where $\qG_{lk}$ is a diagonal matrix whose $(l+(k-1)L)$-th main diagonal element is 1 and 0 elsewhere, $\bar{\qD}=\bar{\qd}\bar{\qd}^T\in\bbC^{KL\times KL}$  with $\bar{\qd}=[\bar{\qd}^T_1,\ldots,\bar{\qd}^T_K]^T$, $\qF_k\in\bbC^{KL\times KL}$ is a matrix whose main diagonal elements from $((k-1)L+1)$-th to $(kL)$-th are 1's and elsewhere 0, $\tilde{\qD}_{ik}=\diag([\qz^T_1,\ldots,\qz^T_{i-1},(\bar{\qd}_i\circ\bar{\qd}_k)^T,\qz^T_{i+1},\ldots,\qz^T_K]^T)$, and
$\qD_k=\text{diag}(\{\qd_{lk}\}^L_{l=1})$ with $\qd_{lk}=[d_{l,k},\ldots,d_{l,k}]^T\in\bbC^{N\times 1}$.

As a result, $\mathcal{P}_{1\text{-}1}$ can be  reformulated as a semidefinite programming as follows:
\begin{subequations}
\begin{align}
  \mathcal{P}_{1\text{-}2}:\min_{\qW,\bm{\Psi},\bm{\Gamma}}  \ \
  &\sum_{l\in\calN_R}\bar{P}_{R_l}+\eta_l\tilde{\bar{R}}_{F_l} \\
   \text{s.t.}\ \ &\tilde{\bar{R}}_{F_l}\leq C_l, \forall l\in\mathcal{N}_R, \label{P12 cond1}\\
  &\qW\succeq0, \label{P12 cond2}\\
  &\text{rank}(\qW)=1, \label{P12 cond3}\\
  &\eqref{P11 cond1} \ \text{and}\ \eqref{P11 cond3},
\end{align}
\end{subequations}
where $\bm{\Gamma}$ is a set of $\bm{\Gamma}_l$'s. The optimal value of  $\bm{\Gamma}_{l}$ in  $\mathcal{P}_{1\text{-}2}$, according to \textbf{Lemma \ref{lemma1}}, is $\bm{\Gamma}^{\ast}_{l}=\bm{\Lambda}^{-1}_l,  \forall l\in\calN_R$. Relaxing the rank constraint $ \text{rank}(\qW)=1$, $\mathcal{P}_{1\text{-}2}$ is still non-convex over three variables $\qW$, $\bm{\Psi}$, and $\bm{\Gamma}$.  But it is convex with respect to any one of these variables  and can converge to a stationary point by an iterative coordinate descent algorithm as shown in \textbf{Algorithm \ref{iterative coordinate descent algorithm}}.

In \textbf{Algorithm \ref{iterative coordinate descent algorithm}}, at $t$ iteration, $\qW^{(t)}$ and $\bm{\Psi}^{(t)}$ are optimized simultaneously, whereas $\bm{\Gamma}^{(t)}$ is updated directly as $\bm{\Gamma}^{(t)}_l=(\bm{\Lambda}^{(t)}_l)^{-1},\forall l\in\calN_R,$ according to  \eqref{inequality1}. Such process is repeated until convergence.  Note that \textbf{Algorithm \ref{iterative coordinate descent algorithm}} does not take the rank-one constraint into consideration. After the semidefinite relaxation (SDR) of $\mathcal{P}_{1\text{-}2}$ is solved, the optimal solution $\qW^{\ast}$ should be converted into a feasible solution to  $\mathcal{P}_{1}$. Since the rank of $\qW^{\ast}$ may not equal to one, we can extract the feasible solution to  $\mathcal{P}_{1}$ from $\qW^{\ast}$ with Gaussian randomization method \cite{luo2010semidefinite}.  \textbf{Algorithm \ref{iterative coordinate descent algorithm}} generates a non-increasing sequence of objective values, thus the convergence is guaranteed  \cite{li2013transmit}.  The main computational complexity of \textbf{Algorithm \ref{iterative coordinate descent algorithm}} lies in step 2, where the SDR of  $\mathcal{P}_{1\text{-}2}$ is solved. The computational complexity of the SDR of $\mathcal{P}_{1\text{-}2}$  is $\mathcal{O}(D^{3.5}_{SDP}\log(1/\epsilon))$ with  a custom-built interior-point algorithm \cite{helmberg1996interior}, where $\epsilon>0$ is the solution accuracy and $D_{SDP}=KL+NL$ is the dimension. Assuming that \textbf{Algorithm \ref{iterative coordinate descent algorithm}} converges in $T_1$ iterations,  the total complexity of  \textbf{Algorithm \ref{iterative coordinate descent algorithm}} is $\mathcal{O}(T_1D^{3.5}_{SDP}\log(1/\epsilon_2))$ \cite{dai2016energy}.
\begin{algorithm}[htb]
\caption{Iterative coordinate descent algorithm.}
\label{iterative coordinate descent algorithm}
\begin{algorithmic}[1]
\STATE \textbf{Initialization:} $\qW^{(0)}$,$\bm{\Psi}^{(0)}_l$,  $\bm{\Gamma}^{(0)}_l=(\bm{\Lambda}^{(0)}_l)^{-1}$ and $t=0$. \\
\STATE Update $t=t+1$ and find the optimal $\bm{\Psi}^{(t)}_l$ and $\qW^{(t)}$ with given $\bm{\Gamma}^{(t-1)}_l$ via solving problem $\mathcal{P}_{1\text{-}2}$.\\
\STATE Update $\bm{\Gamma}^{(t)}_l=(\bm{\Lambda}^{(t)}_l)^{-1}$.\\
\STATE Repeat steps 2 and 3 until convergence.
\STATE Return $\bm{\Psi}^{(t)}_l$ and $\qW^{(t)}$ as the optimal solution $\bm{\Psi}^{\ast}_l$ and $\qW^{\ast}$, respectively.
\end{algorithmic}
\end{algorithm}

\section{Joint Network Power Minimization Problem for Computation and Transmission}\label{Power Minimization Problem for Computation and Transmission}
In this section, we find the  solution to the joint network power minimization problem  $\mathcal{P}_{2}$.
\subsection{Problem Reformulation}
We find that $\mathcal{P}_{2}$ has to confront with  all the difficulties in $\mathcal{P}_{0}$ and $\mathcal{P}_{1}$  because  $\mathcal{P}_{2}$ is combination of  two problems coupled by the delay constraint. To avoid the nonconvexity, we first reformulate  $\mathcal{P}_{2}$  as
\begin{subequations}
\begin{align}
  \mathcal{P}_{2\text{-}1}:\min_{\qx,\qA,\qP,\bm{\Psi}}  \ \ &\sum_{s\in\calN_S}\sum_{k\in\calN_U}\chi_{s,k}A_{s,k}+\omega\sum_{l\in\calN_R}(P_{R_l}+P_{F_l})\\
  \text{s.t.}\ \ &\frac{L_k}{\sum_{s\in\calN_S}\varsigma_{s,k}A_{s,k}}+T^{(TR)}_k\leq \tau_{k}, \forall k\in\mathcal{N}_U, \\
   &\sum_{k\in\calN_U}A_{s,k}\leq \lambda_s,\forall s\in\calN_S,\label{relax1} \\
  &A_{s,k}\leq x_{s,k}\lambda_s,\forall s\in\calN_S,\forall k\in\calN_U, \label{relax2}\\
  &\eqref{P0 cond2}-\eqref{P0 cond5},\eqref{P1 cond2}, \ \text{and} \ \eqref{P1 cond3},
\end{align}
\end{subequations}
where  \eqref{relax2} indicates that  server $s$ does not allocate any resource to task $\Phi_k$, if task $\Phi_k$ is not assigned on server $s$, i.e., $x_{s,k}=0\Longrightarrow A_{s,k}=0$.   As mentioned above,  $\mathcal{P}_{2\text{-}1}$ is a mixed time-scale problem. Similar to  $\mathcal{P}_{1}$, we turn  $\mathcal{P}_{2\text{-}1}$ into a slow time-scale  problem based on the asymptotic results in Section \ref{Power Minimization  Problem for Transmission} and formulate the SDR of $\mathcal{P}_{2\text{-}1}$ as follows:
\begin{subequations}
\begin{align}
  \mathcal{P}_{2\text{-}2}:\min_{\qx,\qA,\qW,\bm{\Psi},\bm{\Gamma},\bm{\varphi}}  \ \ &\sum_{s\in\calN_S}\sum_{k\in\calN_U}\chi_{s,k}A_{s,k}+\omega\sum_{l\in\calN_R}\bar{P}_{R_l}+\eta_l\tilde{\bar{R}}_{F_l} \\
  \text{s.t.}\ \ &\frac{L_k}{\sum_{s\in\calN_S}\varsigma_{s,k}A_{s,k}}+\frac{D_k}{\tilde{\bar{R}}_{U_k}}\leq \tau_{k}, \forall l\in\mathcal{N}_R, \label{p22 cond1}\\
  &x_{s,k}\in[0,1],\label{p22 cond2}\\
  &\eqref{P0 cond3},\eqref{P0 cond4}, \eqref{P11 cond3}, \eqref{P12 cond1}, \eqref{P12 cond2}, \eqref{relax1}, \text{and} \ \eqref{relax2},
\end{align}
\end{subequations}
where $\bm{\varphi}$ is a set of $\varphi_k$'s and
\begin{equation}
\tilde{\bar{R}}_{U_k}=\log_2(\overline{\text{Sig}}_k+\overline{\text{Int}}_k)+\log_2(\varphi_k)-\varphi_k \overline{\text{Int}}_k+1.
\end{equation}

In $\mathcal{P}_{2\text{-}2}$, $x_{s,k}$'s are  relaxed as continuous variables within $[0,1]$ and   \eqref{inequality1} and \eqref{inequality2} are applied to $\bar{R}_{F_l}$  and $\bar{R}_{U_k}$, respectively. Then,  $\mathcal{P}_{2\text{-}2}$ is convex with respect to either  $\{\qx,\qA,\qW,\bm{\Psi}\}$ or $\{\bm{\Gamma},\bm{\varphi}\}$.  Thus, we find the solution to $\mathcal{P}_{2\text{-}2}$ by alternatingly solving the following  two problems:
\begin{subequations}
\begin{align}
  \mathcal{P}_{2\text{-}3}:\mathop{\min}_{\qx,\qA,\qW,\bm{\Psi}}  \ \ &\sum\limits_{s\in\calN_S}\sum\limits_{k\in\calN_U}\chi_{s,k}A_{s,k}+\omega\sum\limits_{l\in\calN_R}\bar{P}_{R_l}+\eta_l\tilde{\bar{R}}_{F_l} \\
 \text{s.t.}\ \ &\eqref{P0 cond3}, \eqref{P0 cond4}, \eqref{P11 cond3}, \eqref{P12 cond1}, \eqref{P12 cond2}, \eqref{relax1},\eqref{relax2},  \eqref{p22 cond1},\text{and} \ \eqref{p22 cond2},
\end{align}
\end{subequations}
and
\begin{subequations}
\begin{align}
  \mathcal{P}_{2\text{-}4}:\mathop{\min}_{\bm{\Gamma},\bm{\varphi}}  \ \ &\sum\limits_{l\in\calN_R}\bar{P}_{R_l}+\eta_l\tilde{\bar{R}}_{F_l} \\
  \text{s.t.}\ \ &\eqref{P11 cond3}, \eqref{P12 cond1},  \ \text{and} \ \eqref{p22 cond1},
\end{align}
\end{subequations}
where the optimal solution to  $\mathcal{P}_{2\text{-}4}$ is given as
\begin{equation}
  \bm{\Gamma}^{\ast}_l=\bm{\Lambda}_l^{-1} \ \text{and} \ \varphi^{\ast}_k=\text{Int}_k^{-1}.
\end{equation}
By applying the dual decomposition  to  $\mathcal{P}_{2\text{-}3}$ \cite{ye2013user,shen2014distributed,Wang2016Joint}, the Lagrangian function associated with problem $\mathcal{P}_{2\text{-}3}$ is given by
\begin{align}
  &L(\qx,\qA,\qW,\bm{\Psi},\bm{\mu})\nonumber\\
  &=\!\sum_{s\in\calN_S}\sum_{k\in\calN_U}\chi_{s,k}A_{s,k}\!+\!\omega\sum_{l\in\calN_R}(\bar{P}_{R_l}\!+\!\eta_l\tilde{\bar{R}}_{F_l})\!+\!\sum_{k\in\calN_U}\mu_k(\frac{L_k}{\sum_{s\in\calN_S}\varsigma_{s,k}A_{s,k}}\!+\!\frac{D_k}{\tilde{\bar{R}}_{U_k}} -\tau_{k}),\nonumber
\end{align}
where  $\bm{\mu}=[\mu_1,\ldots,\mu_K]^T\in\bbC^{K\times 1}$ is composed of the Lagrangian multipliers. The corresponding Lagrangian dual function is given by
\begin{equation}
  g(\bm{\mu})=g_1(\bm{\mu})+g_2(\bm{\mu})-\sum_{k\in\calN_U}\mu_k\tau_{k},
\end{equation}
where
\begin{equation}\label{dual pro2}
\begin{cases}
  g_1(\bm{\mu})=&\inf\limits_{\qx,\qA} \sum\limits_{s\in\calN_S}\sum\limits_{k\in\calN_U}\chi_{s,k}A_{s,k}+\sum\limits_{k\in\calN_U}\mu_k\frac{L_k}{\sum_{s\in\calN_S}\varsigma_{s,k}A_{s,k}},\\
  &\text{s.t.} \ \eqref{P0 cond3},\eqref{P0 cond4}, \eqref{relax1},\eqref{relax2}, \text{and} \ \eqref{p22 cond2},
\end{cases}
\end{equation}
and
\begin{equation}\label{dual pro3}
\begin{cases}
  g_2(\bm{\mu})=&\inf\limits_{\bm{\Psi}, \bm{\qW}} \sum\limits_{l\in\calN_R}(\omega\bar{P}_{R_l}+\omega\eta_l\tilde{\bar{R}}_{F_l})+\sum\limits_{k\in\calN_U}\mu_k\frac{D_k}{\tilde{\bar{R}}_{U_k}},\\
  &\text{s.t.}\ \eqref{P11 cond3}, \eqref{P12 cond1}, \text{and} \ \eqref{P12 cond2}.
\end{cases}
\end{equation}
Then,  the  master dual problem associated with $\mathcal{P}_{2\text{-}3}$ is formulated as
\begin{equation}
   \mathcal{P}_{2\text{-}5:}\mathop{\max}_{\bm{\mu}}g(\bm{\mu}).
\end{equation}
Since $\mathcal{P}_{2\text{-}3}$ is convex and satisfies the Slater's condition, the duality gap of  $\mathcal{P}_{2\text{-}3}$ and its dual problem $\mathcal{P}_{2\text{-}5}$ is zero \cite{boyd2004convex}. In the following, we propose a distributed algorithm based on hierarchical decomposition to find the optimal solution to $\mathcal{P}_{2\text{-}2}$.

\subsection{Distributed Algorithm Based on Hierarchical Decomposition}

In \textbf{Algorithm \ref{Distribued algorithm}}, the upper level primal decomposition is conducted, which introduces  $\mathcal{P}_{2\text{-}3}$ and $\mathcal{P}_{2\text{-}4}$. Based on $\mathcal{P}_{2\text{-}3}$, the lower level dual decomposition is conducted to formulate the dual problem $\mathcal{P}_{2\text{-}5}$. Therefore, the distributed algorithm should involve two level iterations: the outer iteration is for $\mathcal{P}_{2\text{-}3}$ and $\mathcal{P}_{2\text{-}4}$ to converge and the inner iteration is for $\mathcal{P}_{2\text{-}5}$ to converge.

  In the outer iteration, the optimal solution to  $\mathcal{P}_{2\text{-}4}$ is directly given as
$\bm{\Gamma}^{(t)}_l=(\bm{\Lambda}^{(t)}_l)^{-1}$ and $\varphi^{(t)}_k=(\text{Int}^{(t)}_k)^{-1}$. However, to obtain  the optimal solution to $\mathcal{P}_{2\text{-}3}$, it relies on the dual problem $\mathcal{P}_{2\text{-}5}$, whose optimal solution can be achieved via the inner iteration where $(\qx^{(q)}$,$\qA^{(q)})$ and $(\qW^{(q)},\bm{\Psi}^{(q)})$ are alternatingly updated. Specifically, at $p$-th inner iteration:
\begin{enumerate}
  \item \textbf{Data center's algorithm} jointly optimizes task scheduling and computation resource allocation. $\qx^{(p)}$ and $\qA^{(p)}$ are updated by solving subproblem $g_1(\bm{\mu})$ with a BnB algorithm similar to \textbf{Algorithm \ref{BnB algorithm}} or a combinational algorithm similar to \textbf{Algorithm \ref{combinational algorithm}}.
  \item \textbf{BBU pool's algorithm} jointly optimizes power allocation and compression noise. $\qW^{(p)}$ and $\bm{\Psi}^{(p)}$ are updated by solving subproblem $g_2(\bm{\mu})$ with an iterative coordinate descent algorithm similar to \textbf{Algorithm \ref{iterative coordinate descent algorithm}}.

\item  On the other hand, the price factor is adjusted by \textbf{UEs' algorithm}. Since $g(\bm{\mu})$ is not differentiable over $\mu_k$, a sub-gradient approach is adopted here to update the price factor $\mu_k$ at UE $k$, i.e.,
\begin{equation}\label{multipler1}
\mu^{(p+1)}_k=\left[\mu^{(p)}_k+\delta^{(p)}_{\mu}\left(\frac{L_k}{\sum_{s\in\calN_S}\varsigma_{s,k}A^{(p)}_{s,k}}+\frac{D_k}{\tilde{\bar{R}}_{U_k}} -\tau_{k}\right)\right]^{+}, \forall k\in\calN_U,
\end{equation}
\end{enumerate}
where $\delta^{(p)}_{\mu}$ is dynamically chosen stepsize sequence \cite{bertsekas2009convex,ye2013user}. Similarly, after the SDR problem of $\mathcal{P}_{2\text{-}2}$ is solved, we need to extract  $p_{l,k}$'s from $\qW^{\ast}$ with Gaussian randomization method \cite{luo2010semidefinite}. Note that there are three sub-algorithms named data center's algorithm, BBU pool's algorithm, and UEs' algorithm in \textbf{Algorithm \ref{Distribued algorithm}} and they are executed in parallel in the data center, the BBU pool, and the UEs, respectively. Therefore, the complexity is significantly reduced compared to the direct optimization of  $\mathcal{P}_{2\text{-}2}$.

\begin{algorithm}[htb]
\caption{Distributed algorithm based on hierarchical decomposition.}
\label{Distribued algorithm}
\begin{algorithmic}[1]
\STATE \textbf{Initialization:} $\underline{\qW}^{(0)}$, $\underline{\bm{\Psi}}^{(0)}$, $t=0$.  \\
\WHILE {Convergence of outer iteration ($t$) not achieved}
\STATE $t=t+1$ and $p=0$.
\WHILE{Convergence of inner iteration ($p$) not achieved}
\STATE $p=p+1$;
\STATE \textbf{Data center's algorithm:}  Update $\qx^{(p)}$ and $\qA^{(p)}$ by solving subproblem $g_1(\bm{\mu})$.
\STATE \textbf{BBU pool's algorithm:}  Update $\qW^{(p)}$ and $\bm{\Psi}^{(p)}$ by solving subproblem $g_2(\bm{\mu})$.
\STATE \textbf{UEs' algorithm:} Update $\bm{\mu}^{(p)}$ according to \eqref{multipler1}.
\ENDWHILE
\STATE Update $\{\underline{\qx}^{(t)}, \underline{\qA}^{(t)}, \underline{\qW}^{(t)}, \underline{\bm{\Psi}}^{(t)}\}$ as $\{\qx^{(p)}, \qA^{(p)},\qW^{(p)},\bm{\Psi}^{(p)}\}$ at the convergence of the inner iteration.
\STATE Update  $\underline{\bm{\Gamma}}^{(t)}_l=\bm{\Lambda}^{-1}_l(\underline{\qW}^{(t)}, \underline{\bm{\Psi}}^{(t)}), \forall l\in\calN_R,$ and $\underline{\varphi}^{(t)}_k=\text{Int}^{-1}_k(\underline{\qW}^{(t)}, \underline{\bm{\Psi}}^{(t)}), \forall k\in\calN_U$.
\ENDWHILE
\STATE Return $\{\underline{\qx}^{(t)}, \underline{\qA}^{(t)}, \underline{\qW}^{(t)}, \underline{\bm{\Psi}}^{(t)}\}$ as the optimal solution at the convergence of the outer iteration.
\end{algorithmic}
\end{algorithm}

\section{Numerical Results}\label{section numerical results}
In this section, we present the  numerical results to show the performance of our proposed algorithms,  where $L$ RRHs and $K$ UEs are distributed uniformly and independently in an area with a radius  of 100 m. The outer interference combined with background noise is set as -150 dBm/Hz and the path loss function is given as $128.1+37.6\log_{10}(d)$ where $d$ in km. The system bandwidth is $B=20$ MHz and the number of RRH antenna is $N=5$. For simplicity, we assume that each RRH has the same parameters and is subject to the same constraints, i.e., $P^{(MAX)}_{R_l}=1$ W, $C_l=2$ bps/Hz, $\eta_l=0.5$, $\upsilon_l=0.25$, $p^{(Active)}_{R_l}=6.8$ W, and $p^{(Sleep)}_{R_l}=4.3$ W, $\forall l\in\calN_R$. The task load $L_k$ is assumed to be uniformly distributed in $[0.01,0.1]$ and the output data is $D_k=1.6$ Mbits, $\forall k\in \calN_U$. The parameters of the servers are set as  $P^{(Static)}_{S_s}=2$ W, $\chi_{s,k}=1,\forall s\in\calN_S,\forall k\in\calN_U$, and the computing capacity $\lambda_s$ is uniformly distributed in $[\lambda_{lb},\lambda_{ub}]$. Moreover, we assume the execution efficiency $\varsigma_{s,k}$ is distributed uniformly in  $[\varsigma_{lb}, \varsigma_{ub}]$, $c_1=1$, $c_2=10^{-5}$, and $\omega=1$.

\subsection{Power Consumption for Computation}

\begin{figure}
\includegraphics[width=0.75\textwidth]{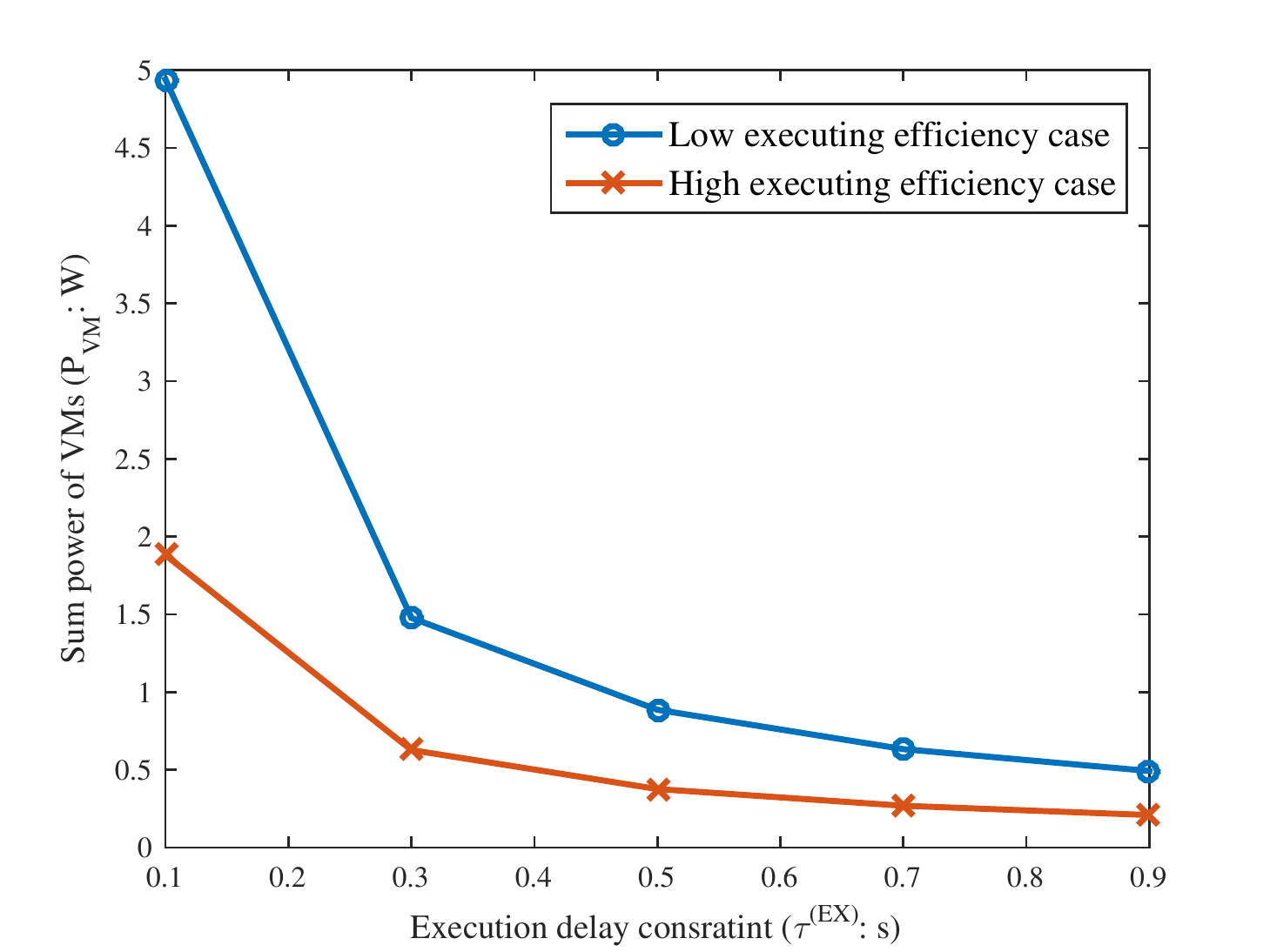}
\centering
\caption{Sum of power consumed by VMs versus execution delay constraint.} \label{change_execution_efficiency}
\end{figure}

We first consider that each task has the same execution delay constraint, i.e., $\tau^{(EX)}_k=\tau^{(EX)},\forall k\in\calN_U$. Fig. \ref{change_execution_efficiency} shows the sum of power consumed by all the VMs, i.e., $P_{VM}=\sum_{s\in\calN_S}\sum_{k\in\calN_U}P_{VM_{s,k}}$, versus the execution delay constraint $\tau^{(EX)}$  with $\{K=6,S=4,\lambda_{lb}=1,\lambda_{ub}=2\}$. It is observed that with the increase of $\tau^{(EX)}$, the consumed power decreases accordingly because the VMs have more time to finish the tasks with a lower power. To study the influence of the execution efficiency $\varsigma_{s,k}$ on the power consumption of the VMs, we set two different regimes $[0.1,0.5]$ and [0.6,1] representing low and high execution efficiency cases, respectively. It is observed that  higher execution efficiency leads to less power consumption.

\begin{figure}
  \centering
  \subfigure[]{
    \label{power_comp_greedy_bnb} 
    \includegraphics[width=0.75\textwidth]{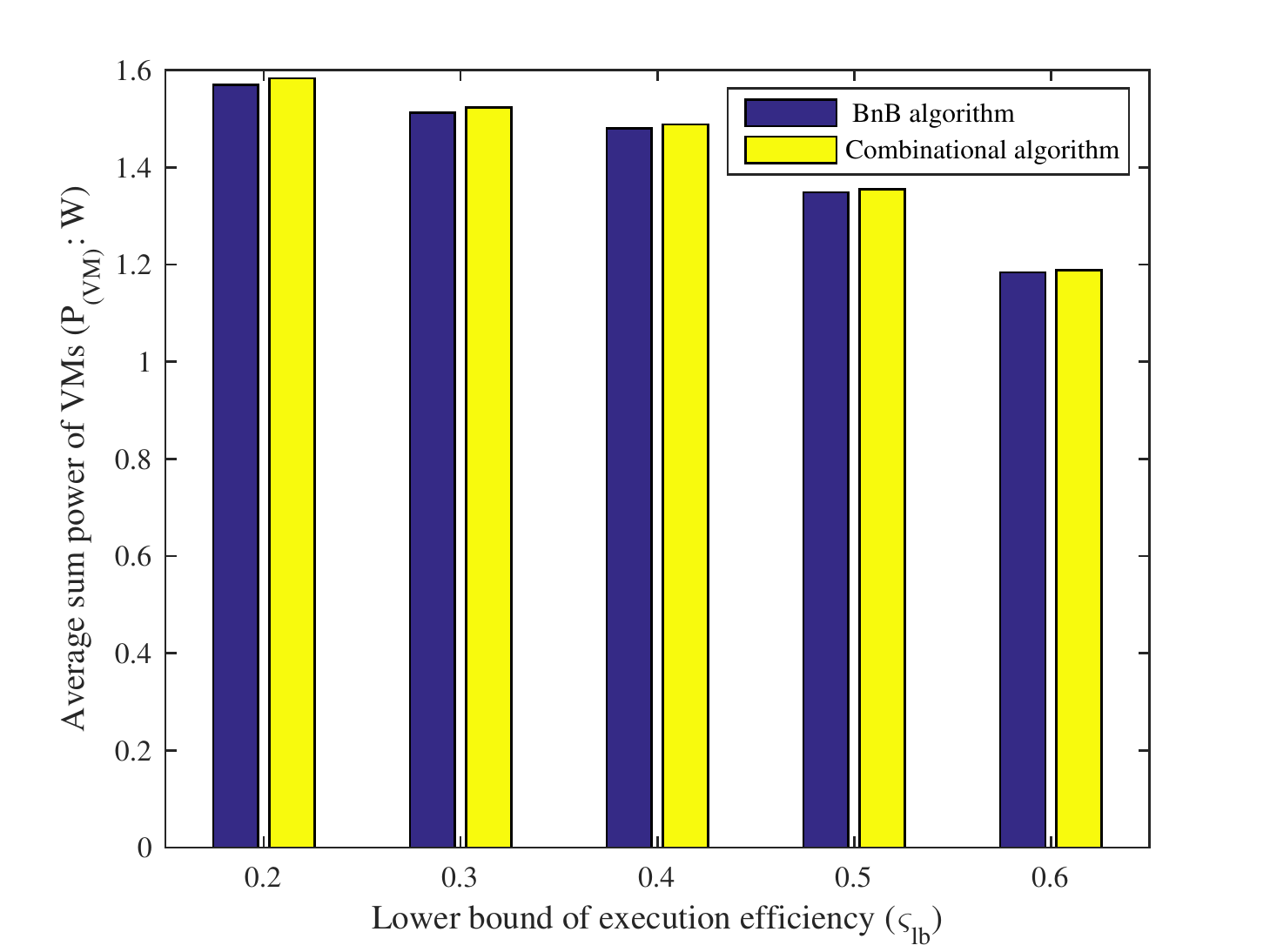}}
  \hspace{1in}
  \subfigure[]{
    \label{runtime_comp_greedy_bnb} 
    \includegraphics[width=0.75\textwidth]{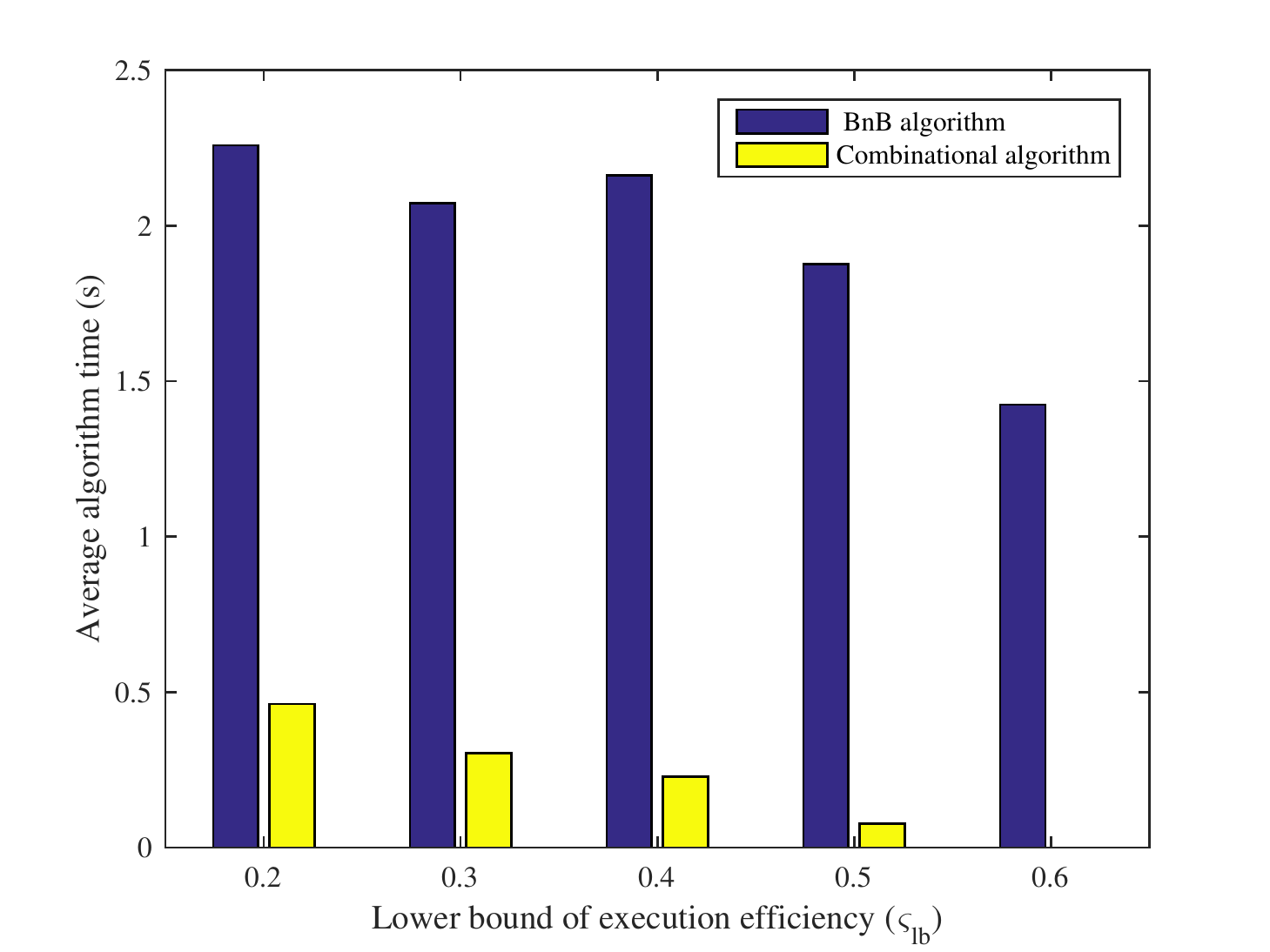}}
  \caption{Comparison of BnB algorithm and combinational algorithm: (a) average power consumption of the VMs and (b) average algorithm time.}
 \label{comp_greedy_bnb} 
\end{figure}

Next, we compare the BnB algorithm (i.e., \textbf{Algorithm \ref{BnB algorithm}}) with the combinational algorithm (i.e., \textbf{Algorithm \ref{combinational algorithm}}) versus the execution efficiency $\varsigma_{s,k}$ in Fig. \ref{comp_greedy_bnb} with $\{K=6,S=4,\lambda_{lb}=0.1,\lambda_{ub}=1\}$. Since $\varsigma_{s,k}$'s are random variables distributed uniformly in $[\varsigma_{lb},\varsigma_{ub}]$, we divide the value range into many segments $[\varsigma_{lb},\varsigma_{lb}+0.1] $ with a fixed length 0.1 for fairness  and use the lower bound of the execution efficiency $\varsigma_{lb}$ as the x-axis. Fig. \ref{comp_greedy_bnb} is the average result of 200 independent realizations.  It is found that the power consumption for computation decreases as the lower bound of the execution efficiency $\varsigma_{lb}$ increases because the demand for computation resource is reduced. We also observe that the solutions obtained by the combinational task scheduling algorithm are suboptimal and require a little more power consumption but with much less runtime, compared to the BnB algorithm.  Therefore, in order to save time and reduce computation complexity, it is suggested to adopt the combinational task scheduling algorithm with a little performance loss. However, the optimal solution can be found via the BnB algorithm  at the cost of computational time and complexity. Besides, Fig. \ref{runtime_comp_greedy_bnb} suggests that as the lower bound of the execution efficiency $\varsigma_{lb}$ increases, indicating that the overall execution efficiency of servers is improved, then the fraction of times where \textbf{Algorithms 2} fails decreases.

\subsection{Power Consumption for Transmission}
\begin{figure}
\includegraphics[width=0.75\textwidth]{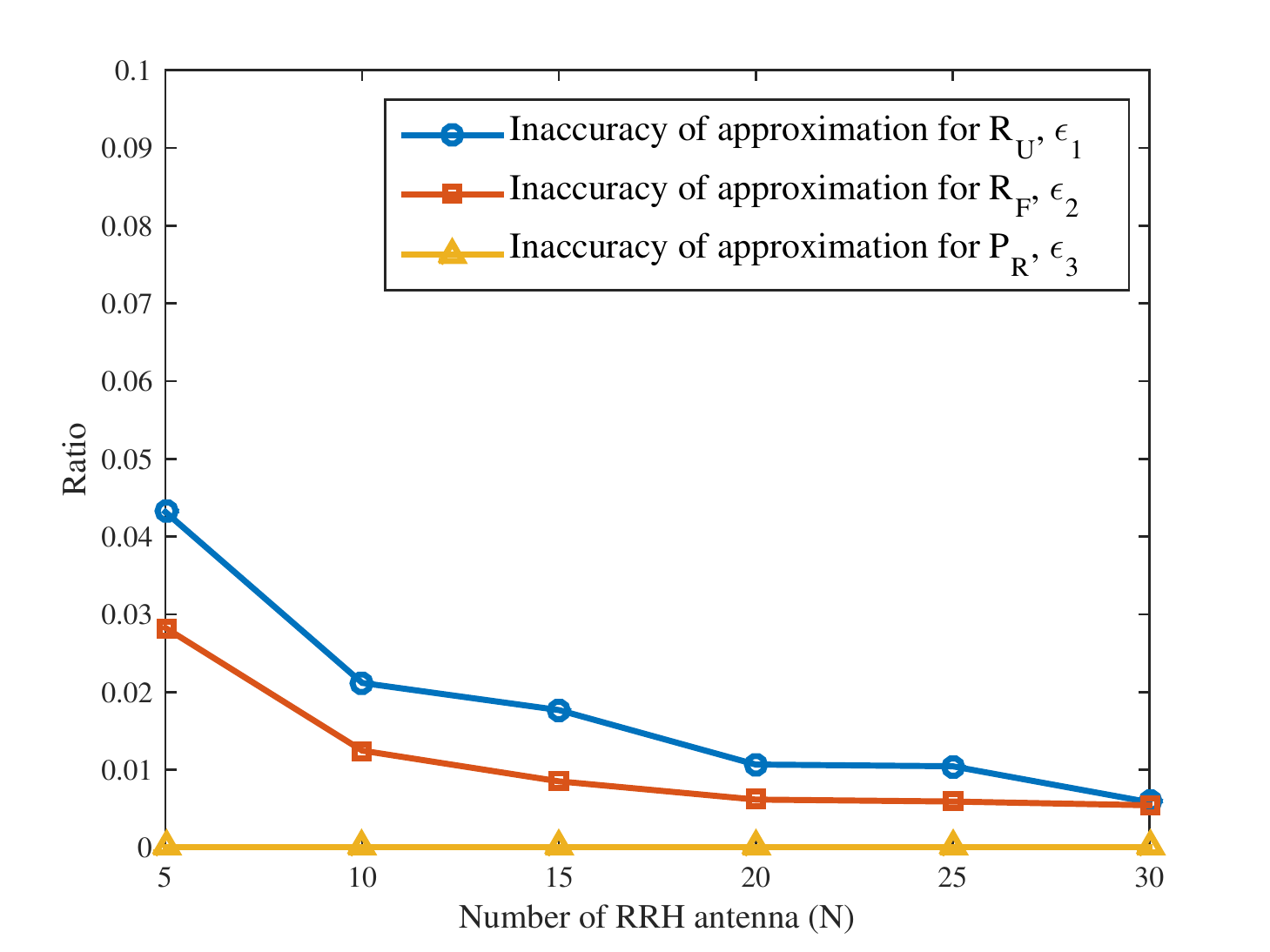}
\centering
\caption{Accuracy of approximative results.} \label{inaccuracy_of_approximation}
\end{figure}

In the following, we first validate the accuracy of the approximate results derived in Section \ref{Power Minimization  Problem for Transmission}. We define $\epsilon_1=\sum_{k\in\calN_U}\frac{|\bar{R}_{U_k}-R_{U_k}|}{R_{U_k}}$, $\epsilon_2=\sum_{l\in\calN_R}\frac{|\bar{R}_{F_l}-R_{F_l}|}{R_{F_l}}$, and $\epsilon_3=\sum_{l\in\calN_R}\frac{|\bar{P}_{R_l}-P_{R_l}|}{P_{R_l}}$ as the inaccuracy levels of the sum rate of UEs $\bar{R}_{U}=\sum_{k\in\calN_U}\bar{R}_{U_k}$, the sum rate of fronthaul links $\bar{R}_{F}=\sum_{l\in\calN_R}\bar{R}_{F_l}$, and the sum power of RRHs $\bar{P}_{R}=\sum_{l\in\calN_R}\bar{P}_{R_l}$, respectively. Fig. \ref{inaccuracy_of_approximation} shows that these approximate results are not only  close to their original expressions but  become more accurate as the number of RRH antennas $N$ increases.

\begin{figure}
\includegraphics[width=0.75\textwidth]{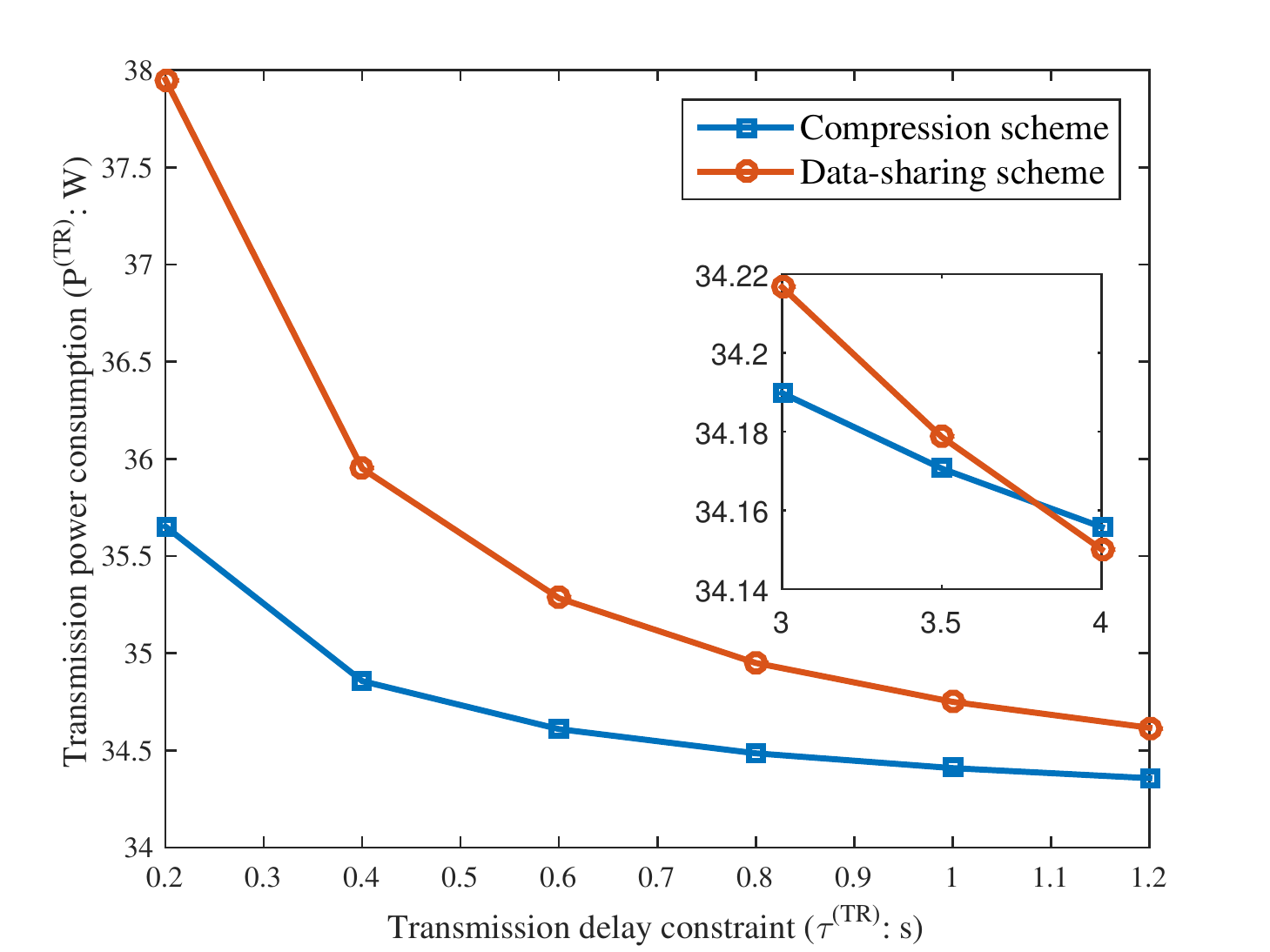}
\centering
\caption{Comparison of compression and data-sharing strategies.} \label{comp_datasharing_compress}
\end{figure}

For notational simplicity,  we define $P^{(TR)}=\sum_{l\in\calN_R}(P_{R_l}+P_{F_l})$ as the total power consumption for transmission. It is also assumed that each UE has the same transmission delay, i.e., $\tau^{(TR)}_k=\tau^{(TR)},\forall k\in\calN_U$.  To compare compression-based transmission scheme with the data-sharing transmission scheme, Fig. \ref{comp_datasharing_compress} plots  $P^{(TR)}$ versus the transmission delay constraint $\tau^{(TR)}$.  It can be  observed that the power consumption of both transmission schemes decrease as the transmission delay constraint increases. In addition, Fig. \ref{comp_datasharing_compress} indicates that with  strict transmission delay constraint, i.e., small values of $\tau^{(TR)}$, compression-based transmission scheme produces less power consumption. However, when the  transmission delay constraint is loose, i.e., large values of $\tau^{(TR)}$, the data-sharing transmission scheme achieves a better performance. This is because the fronthaul rate of compression scheme relies on the signal-to-quantization-noise ratio whereas that of  data-sharing scheme depends on the UEs' rates and the serving RRH numbers, since the data-sharing scheme delivers each UE's message to all the RRHs that serve this UE via fronthaul links. A smaller value of $\tau^{(TR)}$ suggests a higher data-rate demand, then  more RRHs are required to serve the UEs. Therefore, a faster increase of the fronthaul rate occurs in the data-sharing scheme. However, a gradual increase of the fronthaul rate in the compression scheme as the data-rate demand rises.

\subsection{Joint Network Power Minimization Problem for Computation and Transmission}
\begin{figure}
\includegraphics[width=0.75\textwidth]{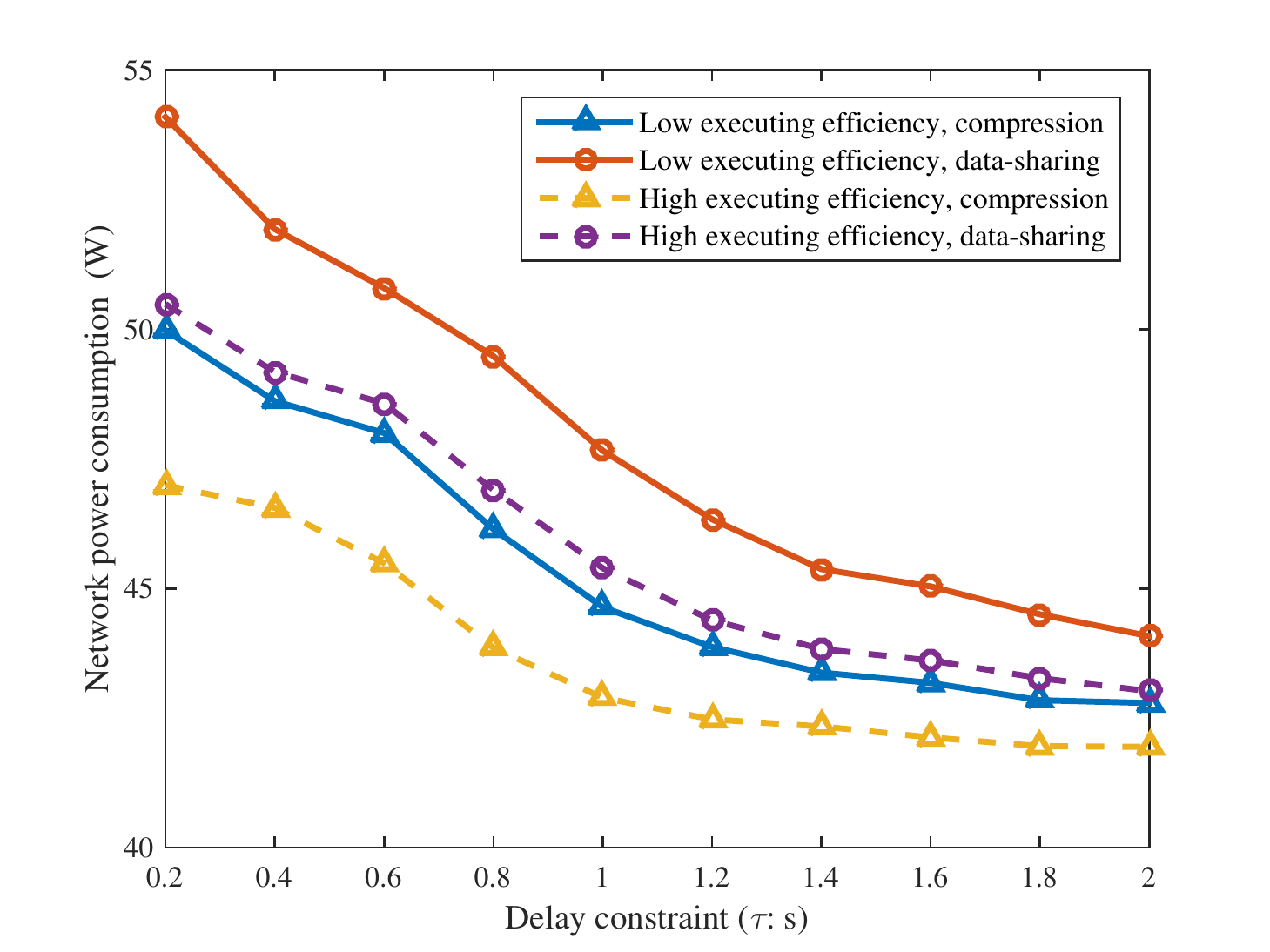}
\centering
\caption{Network power consumption versus different delay constraint $\tau$ values under different transmission schemes and different executing efficiency cases.} \label{npc}
\end{figure}

Finally, we present the network power minimization with respect to the delay constraint $\tau_k$ under different transmission schemes (i.e., compression based scheme and data-sharing based scheme) and different executing efficiency cases (i.e., low executing efficiency case with $\{\varsigma_{lb}=0.1,\varsigma_{ub}=0.5\}$ and high executing efficiency case with $\{\varsigma_{lb}=0.6,\varsigma_{ub}=1\}$) in Fig. \ref{npc}. For simplicity, we assume $\tau_k=\tau,\forall k\in\calN_U$. The network power consumption decreases with the increase of the delay constraint because when the delay constraint increases, the QoS  level decreases and less power is required to meet the QoS. Similarly, when the average executing efficiency is improved, less computational resource is required thus the network power consumption is also reduced. It is also observed  that the network adopts the transmission scheme based on compression shows a better performance than  the data-sharing transmission scheme.
\section{Conclusion}\label{section conclusion}
In this paper, we considered the network power consumption including power consumptions for computation  and transmission in a downlink C-RAN. The power minimization problem for computation  was a slow time-scale problem, since the joint design of task scheduling and computing resource allocation was generally executed much slower than milliseconds. However, the power minimization problem for transmission was a fast time-scale problem because the joint optimization of power allocation and compression was based on  small-scale fading. Therefore, the joint network power  minimization problem was a mixed time-scale problem. To overcome the  time-scale challenge, we introduced the approximate results of the original problems according to large system analysis. The approximate results were dependent on statistical channel information and independent on  small-scale fading, thus the fast/mixed time-scale problem was turned into a slow time-scale one. We proposed a BnB algorithm and a combinational algorithm to find the optimal and suboptimal solutions to the power minimization problem for computation, respectively, and introduced an iterative coordinate descent algorithm to find solutions  to  the power minimization problem for transmission. Then a distributed algorithm based on hierarchical decomposition was also proposed to solve the joint  network power minimization problem. Simulation results showed that for the power minimization problem for computation,  the combinational algorithm achieved the suboptimal solutions with much less computational complexity and time, compared to the BnB algorithm. In addition, as the delay constraint increased, suggesting the decrease of the QoS demand, the joint network power consumption was also reduced.

\appendix
\section*{Proof of \textbf{Lemma} 1}
Based on the law of large numbers, results \eqref{appro 1} and \eqref{appro 2} can be directly obtained with the following expressions \cite{hoydis2013massive,ngo2013energy}:
\begin{equation}
  \frac{1}{N}\tilde{\qh}_{l,k}\tilde{\qh}_{l,k}^\dag\xrightarrow{\ntoinfty} \frac{1}{N}\qI_N\ \text{and}\  \frac{1}{N}\tilde{\qh}_{l,k}\tilde{\qh}_{l,k^{\prime}}^\dag\xrightarrow{\ntoinfty} \qzero, k^{\prime}\neq k.
\end{equation}
Then we focus on result \eqref{appro 3}. We first define  a function $f(z)=\log_2|\qH_{l}\qP_l\qH_{l}^{\dag}+z\qI_N+\bm{\Psi}_l|$, which tends to the numerator of  $R_{F_l}$ as $z\rightarrow0$.  The derivative of $f(z)$ over $z$ is
\begin{equation}
  \frac{\partial f(z)}{\partial z}=\frac{1}{\log2}\tr(\qH_{l}\qP_l\qH_{l}^{\dag}+z\qI_N+\bm{\Psi}_l)^{-1}.
\end{equation}
Using the random matrix theory, we have
\begin{align}\label{trace_eq}
  \tr(\qH_{l}\qP_l\qH_{l}^{\dag}+z\qI_N+\bm{\Psi}_l)^{-1}&\asymp\tr\left(\sum\limits_{k\in\calN_U}{\frac{p_{l,k}d_{l,k}}{1+e_{l,k}}}\qI_N+z\qI_N+\bm{\Psi}_l\right)^{-1}.
\end{align}
Then according to \cite{wen2013adetermini,zhang2013oncap}, we have result \eqref{appro 3} with $z\rightarrow0$.
$\hfill\blacksquare$
\bibliographystyle{IEEEtran}
\bibliography{reference}

\end{document}